\documentclass[preprint]{aastex}

\def\be{\begin{equation}}
\def\ee{\end{equation}}

\begin{document}

\title{Galaxy Number Counts from the Sloan Digital Sky Survey Commissioning Data\footnote{Based on observations obtained with the Sloan Digital Sky Survey.}}

\author{%
Naoki Yasuda\altaffilmark{2},
Masataka Fukugita\altaffilmark{3,4},
Vijay K. Narayanan\altaffilmark{5},
Robert H. Lupton\altaffilmark{5}, 
Iskra Strateva\altaffilmark{5},
Michael A. Strauss\altaffilmark{5},
\v{Z}eljko Ivezi\'{c}\altaffilmark{5},
Rita S.J. Kim\altaffilmark{5},
David W. Hogg\altaffilmark{4}, 
David  H. Weinberg\altaffilmark{6},
Kazuhiro Shimasaku\altaffilmark{7},
Jon Loveday\altaffilmark{8},
James Annis\altaffilmark{9},
Neta A. Bahcall\altaffilmark{5},
Michael Blanton\altaffilmark{9},
Jon Brinkmann\altaffilmark{10},
Robert J. Brunner\altaffilmark{11},
Andrew J. Connolly\altaffilmark{12},
Istv\'{a}n Csabai\altaffilmark{13},
Mamoru Doi\altaffilmark{7},
Masaru Hamabe\altaffilmark{7},
Shin-Ichi Ichikawa\altaffilmark{2}
Takashi Ichikawa\altaffilmark{14},
David E. Johnston\altaffilmark{15},
Gillian R. Knapp\altaffilmark{5},
Peter Z. Kunszt\altaffilmark{13},
Don Q. Lamb\altaffilmark{15},
Timothy A. McKay\altaffilmark{16},
Jeffrey A. Munn\altaffilmark{17},
Robert C. Nichol\altaffilmark{18},
Sadanori Okamura\altaffilmark{7},
Donald P. Schneider\altaffilmark{19},
Gyula P. Szokoly\altaffilmark{20},
Michael S. Vogeley\altaffilmark{21},
Masaru Watanabe\altaffilmark{22}, and
Donald G. York\altaffilmark{15}
}

\altaffiltext{2}{National Astronomical Observatory, Mitaka, Tokyo 181-8588, Japan}
\altaffiltext{3}{Institute for Cosmic Ray Research, University of Tokyo, Kashiwa 277-8582, Japan}
\altaffiltext{4}{Institute for Advanced Study, Olden Lane, Princeton, NJ 08540, U.S.A.}
\altaffiltext{5}{Princeton University Observatory, Princeton, NJ 08544, U.S.A.}
\altaffiltext{6}{Department of Astronomy, The Ohio State University, Columbus, OH 43210. U.S.A.}
\altaffiltext{7}{Department of Astronomy and Research Center for the Early Universe, School of Science, University of Tokyo, Tokyo, 113-0033, Japan}
\altaffiltext{8}{Sussex Astronomy Centre, University of Sussex, Falmer, Brighton BN1 9QJ, UK}
\altaffiltext{9}{Fermi National Accelerator Laboratory, P.O. Box 500, Batavia, IL 60510, U.S.A.}
\altaffiltext{10}{Apache Point Observatory, 2001 Apache Point Road, P.O. Box 59, Sunspot NM 88349-0059, U.S.A.}
\altaffiltext{11}{Department of Astronomy, California Institute of Technology, Pasadena, CA 91125, U.S.A.}
\altaffiltext{12}{University of Pittsburgh, Department of Physics and Astronomy, 3941 O'Hara Street, Pittsburgh, PA 15260, U.S.A.}
\altaffiltext{13}{Department of Physics and Astronomy, The Johns Hopkins University, Baltimore, MD 21218, U.S.A.}
\altaffiltext{14}{Astronomical Institute, Tohoku University, Sendai 980-8578, Japan}
\altaffiltext{15}{Department of Astronomy and Astrophysics, University of Chicago, 5640 South Ellis Avenue, Chicago, IL 60637, U.S.A.}
\altaffiltext{16}{Department of Physics, University of Michigan, 500 East University, Ann Arbor, MI 48109-1120, U.S.A.}
\altaffiltext{17}{U.S. Naval Observatory, Flagstaff Station, P.O. Box 1149, Flagstaff, AZ 86002, U.S.A.}
\altaffiltext{18}{Department of Physics, Carnegie Mellon University, 5000 Forbes Avenue, Pittsburgh, PA 15213-3890, U.S.A.}
\altaffiltext{19}{Department of Astronomy and Astrophysics, 525 Davey Laboratory, Pennsylvania State University, University Park, PA 16802, U.S.A.}
\altaffiltext{20}{Astrophysikalisches Institut Potsdam, Germany}
\altaffiltext{21}{Department of Physics, Drexel University, 3141 Chestnut Street, Philadelphia, PA 19104, U.S.A.}
\altaffiltext{22}{Japan Science and Technology Corporation, Tokyo 102-0081, Japan}

\begin{abstract}

We present bright galaxy number counts in five broad bands
($u',\ g',\ r',\ i',\ z'$) from imaging data taken during the commissioning
phase of the Sloan Digital Sky Survey (SDSS).
The counts are derived from two independent stripes of imaging scans
along the Celestial Equator, one each toward the North and the South 
Galactic cap, covering about 230 and 210 square degrees, respectively.
A careful study is made to verify the reliability of the photometric catalog.
For galaxies brighter than $r^* = 16$ , the catalog 
produced by automated software is examined against eye inspection of all
objects. 
Statistically meaningful results on the galaxy counts are obtained
in the magnitude range $12 \le r^* \le 21$, using a sample of 900,000
galaxies.
The counts from the two stripes differ by about $30\%$ at magnitudes
brighter than $r^*= 15.5$, consistent with a local $2\sigma$
fluctuation due to large scale structure in the galaxy distribution.
The shape of the number counts-magnitude relation brighter than $r^* = 16$
is well characterized by $N \propto 10^{0.6m}$, the relation expected
for a homogeneous galaxy distribution in a ``Euclidean'' universe.
In the magnitude range $ 16 < r^* < 21$, the galaxy counts from both 
stripes agree very well, and follow the prediction of the no-evolution
model, although the data do not exclude a small amount of evolution. 
We use empirically determined color
transformations to derive the galaxy number counts in the $B$ and
$I_{814}$ bands. We compute the luminosity density of the universe at
zero redshift in the five SDSS bands and in the $B$ band. We find 
${\cal L}_{B} = 2.4 \pm 0.4 \times 10^8L_\odot h $Mpc$^{-3}$, 
for a reasonably wide range of parameters of the Schechter 
luminosity function in the $B$ band.

\end{abstract}


\section{Introduction}

The number counts of galaxies as a function of magnitude is one of the 
classical cosmological tests. It has been repeatedly studied over 
many years by many authors, and has yielded valuable insights about
the properties of galaxies and their evolution with redshift
(see \citet{Sandage} and \citet{KooKron} for reviews). However,
despite much effort, the normalization of
the counts-magnitude relation for bright galaxies remains uncertain by
as much as 50\%. 
This leads to uncertainties in the normalization of the galaxy
luminosity function at zero redshift at the same level, and 
therefore to uncertainties in the interpretation of data on 
galaxy evolution with redshift.
The slope of the galaxy counts at bright magnitudes is also a matter of 
debate; some authors have claimed that it is steeper than the slope 
predicted by the
no-evolution model (\citet{Maddox}, but see also 
\citet{weir95}), and argued either for a local 
underdensity in the galaxy distribution or for a rapid evolution of
the galaxy population since $z \approx 0.2$.

Bright galaxies are rare on the sky, and their number density varies due 
to large-scale structure. Hence, we require imaging surveys extending over 
large solid angles to obtain reliable galaxy counts at the bright end.  
For this reason, most bright galaxy counts have been based on photographic 
plate material from Schmidt telescopes \citep{Maddox,BertinDennefeld},
for which it is notoriously difficult to obtain accurate photometric
calibration.
Surveys with CCD imagers \citep{Hall,Tyson,Lilly,Metcalfe} 
have made dramatic progress in 
determining the galaxy counts at faint magnitudes, but the field of view
covered by these surveys is usually small, making it difficult
to accurately measure the number counts at bright magnitudes.
Even the ambitious surveys of \citet{Gardner} and
\citet{Postman} are limited to 8.5 deg$^2$ and 14.7 deg$^2$, 
respectively, much smaller than the fields surveyed
using photographic plates.  What is clearly needed is an
imaging survey of the sky using modern CCD detectors, covering at
least several hundred square degrees. 

The Sloan Digital Sky Survey (SDSS; \citet{York}) consists of an
imaging survey in five photometric bands of $\pi$ steradians of
the Northern sky, as well as a follow-up spectroscopic survey of
roughly $10^{6}$ galaxies and $10^{5}$ quasars, complete within precisely
defined selection criteria. 
Images of the sky are obtained by SDSS at the rate of 20 square degrees 
an hour.
Thus, useful data to determine the galaxy counts at bright magnitudes may be 
obtained in a few nights of imaging.  
In this paper, we present galaxy number counts in the five SDSS passbands 
$u',\ g',\ r',\ i',$ and $z'$, 
using imaging scans  taken during the commissioning phase of the
survey. We derive the galaxy counts in the magnitude range $12.5 < r^* < 21$,
from two patches of the sky along the Celestial Equator, one each
toward the North and the South Galactic cap, covering an area of about 
230 deg$^2$ and 210 deg$^2$, respectively.
We emphasize that the number counts of
galaxies constitute a crucial verification of the photometric catalog 
derived from the imaging survey, in particular on the uniformity of
photometric measurements with magnitude.

The outline of this paper is as follows.
In \S2, we give a brief description of the observations and the data. 
We describe the photometry of galaxies, and the associated errors,
in \S3. In \S4, we describe the construction of the galaxy catalog in
the Northern equatorial stripe, from the full photometric catalog. We
present the  galaxy number counts in the five SDSS bands in \S5. 
In \S6, we convert our number counts to the $B$ and $I_{814}$ bands
using empirical color transformations, and 
compare  them with existing galaxy count results in the literature.
We calculate the luminosity density of the universe in the SDSS bands
and in the $B$ band in \S7. As a consistency test we recalculate
the normalization of the luminosity functions for the SDSS passbands,
which are obtained from the spectroscopic survey of the SDSS project,
and are published prior to this paper \citep{Blanton}.
In \S8, we derive the galaxy counts from the Southern equatorial stripe.
We present the color distributions of galaxies in \S9.
We present our conclusions in \S10.

\section{Imaging Observations and Data}
\label{sec:2}

The SDSS survey is carried out using a wide-field 2.5 m
telescope, 
a large format imaging camera \citep{Gunn}, two fiber-fed
double spectrographs, and an auxiliary 0.5 m telescope for photometric
calibration. 
The sky is imaged in five passbands, $u',\ g',\ r',\ i',\ z'$ 
covering the entire optical
range from the atmospheric ultraviolet cutoff in the blue to the sensitivity 
limit of silicon CCDs in the red \citep{Fukugita:1996}.  The imaging camera 
contains thirty photometric CCDs, arranged in six columns of five rows, each 
row corresponding to a different filter.  The data are taken in
time-delay-and-integrate (TDI, or drift-scan) mode at the sidereal
rate along great circles on the sky, yielding a {\em strip} consisting
of six very long and narrow {\em scanlines}, each $13.5^\prime$ wide.
The effective exposure time (i.e., the transit time of an object
across a single chip) is 54.1 sec.  The scanlines of a given strip do
not overlap, but observing a second strip offset from the first strip
by about $12.8^\prime$ gives
a filled {\em stripe} $2.5^\circ$ wide. 
The scanline in five colors produced by each column of five CCDs is
divided into $2048\times1489$ pixel {\it frames} with an overlap
of 128 pixels between adjacent frames; a set of five frames covering
a given region of sky (in the colors $u'$, $g'$, $r'$, $i'$ and $z'$)
is referred to as a {\it field}.
The pixels are $0.396''$
square on a side, which satisfies the Nyquist limit for seeing 
$>1.0''$ full width at half maximum (FWHM) (see \citet{Gunn} for technical
details). The  0.5m telescope observes standard stars 
to determine the
atmospheric extinction on each night; the zeropoint of the 2.5m
telescope is determined by observing patches of sky that the 2.5m is
observing as well.

The scans of the Northern equatorial stripe were obtained in
two nights of SDSS commissioning observations on March 20 and 21, 1999,
with the telescope parked on the Celestial Equator (SDSS Runs 752 and 756).
The total observation time was about seven hours in Run 752 and about
eight hours in Run 756. The two runs together form a filled stripe. 
For $r' < 16$, we present the galaxy counts from data in the right
ascension range over which a full stripe is completed by this pair of runs: 
$145.^\circ14 < \alpha < 236.^\circ42$ and $-1.^\circ26 < \delta < 1.^\circ26$
(all coordinates are in J2000), covering a total area of  230 deg$^2$. 

The imaging data used in this paper were taken before the commissioning
of the current Photometric Telescope. Therefore, we calibrated the data by
observing secondary patches in the survey area using a (now
decommissioned) 61 cm telescope at the observatory site, and by
observing primary standard stars using the US Naval Observatory's 
40$''$ telescope, with filter and CCD characteristics nominally identical 
to those at the SDSS Photometric Telescope.  Since the transformation from the 
primary standard stars to the objects observed with the SDSS 2.5m telescope 
has not yet been fully defined, we expect photometric errors of
$\approx$0.05 mag in each band with respect to the proper SDSS
photometric system.  
Thus, in this paper, we will denote our magnitudes as 
$u^*$, $g^*$, $r^*$, $i^*$, and $z^*$, to emphasize the preliminary nature 
of our photometric calibration, rather than the notation $u'$, $g'$, $r'$, 
$i'$, and $z'$ that will be used for the final SDSS photometric system.  
However, we will use the latter notation to refer to the SDSS photometric 
passbands themselves. Our magnitude system is based on the AB$_{95}$ system
\citep{Fukugita:1996}. The mean flux density over each broad passband is
$\overline{f_\nu}=3631 \times 10^{-0.4m}$ Jy.

\begin{figure}
\plotone{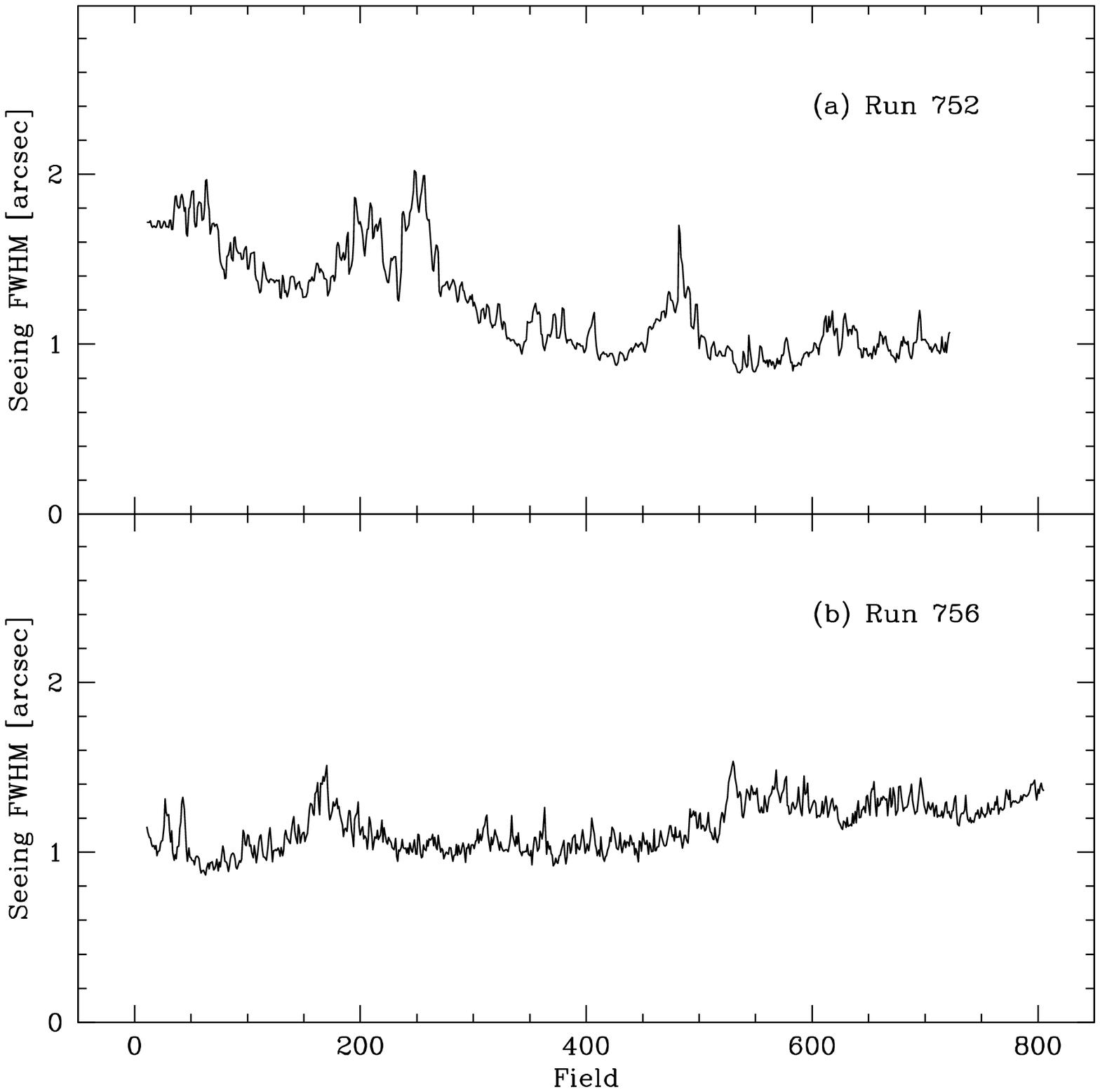}
\caption{Field to field variation of seeing (FWHM) in the $r'$ band in the
third CCD column, during TDI observations: (a) Run 752, extending for 7 hours,
(b) Run 756, extending for 8 hours.}
\label{FWHM}
\end{figure}

Figure \ref{FWHM} shows the FWHM of the point spread 
function (PSF) in the $r'$ band (the seeing) in each field of the third 
CCD column. The PSF is fit to a pair of concentric circular Gaussians; 
the FWHM shown in this
figure is that of this model \footnote{
Note that this FWHM measure gives a value somewhat smaller than that
reported in some other SDSS papers, where the size of seeing is defined
by the square root of the 
effective number of pixels calculated from the ratio of the
first moment square to the second moment of the flux.}. 
The top and bottom panels show the seeing 
variations in Runs 752 and 756, respectively.
The median PSF FWHM in the $r'$ band was $1.4''$ in Run 752, and $1.3''$ in 
Run 756, although the seeing varied substantially over the duration of
the runs.
The sky brightness also varied significantly during the runs. For example
in Run 756 brightness of the $u'$ and $g'$ bands varied by 0.75 mag, that of
$r'$ by 0.45 mag, $i'$ by 0.23 mag and $z'$ by 0.61 mag during the scan
for 8 hours.  For the $z'$ band we recognize a 10\% variation 
in a short time scale ($\approx$ 10 min) 
in addition to a global change.
The average sky brightness was 22.77, 22.06, 21.05, 20.34, and
19.18 mag per square arcsec in $u'$, $g'$, $r'$, $i'$, and $z'$,
respectively.

\section{Photometry of Galaxies}
\label{sec:photometry}

The imaging data are processed with the {\it photometric pipeline} 
(hereafter {\it Photo}; Lupton et al. unpublished, see also \citet{York}
and {\tt http://astro.princeton.edu/PBOOK/welcome.htm})
specifically written for reducing the SDSS data.
Each field is processed independently by {\it Photo}.
Galaxies do not have sharp edges, nor unique profiles of
surface brightness distributions, which
makes it non-trivial to define a
flux for each object. {\it Photo} measures a number of different
types of magnitudes for each object. We present the galaxy counts
using the Petrosian flux \citep{Petrosian}, which is defined by
\begin{equation}
F_P = 2 \pi \int_0^{k r_P} I(r) r dr\ ,
\end{equation}
where $I(r)$ is the surface brightness profile of the object in
question, and $r_P$ is the Petrosian radius satisfying
\begin{equation}
\eta = \frac{I(r_P)}{2 \pi \int_0^{r_P} I(r) r dr / (\pi r_P^2)}\ .
\end{equation}
In practice, this implicit equation for $r_P$ is replaced with
\begin{equation}
\eta = \frac{2 \pi \int_{0.8 r_P}^{1.25 r_P} I(r) r dr / \left[ \pi ((1.25
r_P)^2 - (0.8 r_P)^2)\right]}{2 \pi \int_0^{r_P} I(r) r dr / (\pi r_P^2)}\ ,
\end{equation}
which yields robust measurements. Petrosian radius is independent of the 
distance of a galaxy and foreground reddening. 
Note that two parameters, $k$ and
$\eta$, are needed to specify the Petrosian flux. 
We adopt $k = 2$ and $\eta = 0.2$ to optimize requirements.
A small $\eta$ is preferred in order to minimize the discrepancy in
apertures for galaxies with the exponential profile and those with
the de Vaucouleurs profile. A small $\eta$ also makes the measurement
of the Petrosian radius insensitive to seeing variations.
On the other hand, consideration of signal to noise leads us to 
choose $\eta \geq 0.2$.
 
For every object, {\it Photo}\  calculates the Petrosian radius from 
aperture photometry using spline interpolation.
With our adopted values of $k$ and $\eta$, 
the Petrosian flux is an integration over an aperture of radius
of 3.6 half-light
radii for an object with a de Vaucouleurs profile, and 7.2 scale lengths
for an object with an exponential profile. 
To quantify the difference between Petrosian and total
magnitudes for different galaxy morphologies, we carry out simulations 
with
an empirical model for the mix of galaxy morphology and the scale size
(\citet{Shimasaku}). The source of the other parameters used is explained 
in Section 5 below.
We assume that the de Vaucouleurs profile is truncated at 5 
half light radii \footnote{With the untruncated de Vaucouleurs profile,
11\% of the flux is distributed outside the aperture of this radius. 
The total flux of simulations is renormalized
to the truncated de Vaucouleurs profile. Without the truncation the
tail of the de Vaucouleurs profile causes a 2$-$3\% error in the
local sky estimate.}.
We find that the Petrosian magnitude underestimates the total flux by
$\approx\ 0.03\pm0.01$ mag (mean value) for this morphological mix, 
although the offset can be as 
large as 0.15$-$0.2 mag for face-on de Vaucouleurs profiles.
The Petrosian radius is computed in all color bands, but we measure
the Petrosian flux in any color band using the Petrosian radius 
for the $r'$ band.

We have compared the photometry of objects that lie in the overlap
regions between Runs 752 and 756. We find that the difference between
the magnitudes of the same objects in these two runs taken under
different seeing conditions is consistent with the errors in the
magnitudes quoted by {\it Photo}. At $m = 19$, the errors in Petrosian
magnitudes are about $\pm(0.03 - 0.04)$ mag in the $g', r', i'$ bands and 
about $0.05$ mag in the $u'$ and $z'$ band. 

We have studied the completeness limit of the SDSS imaging data,
by comparison with simulated images, and from comparison 
with deep HST images.  
Lupton et al. (unpublished) present a detailed study of
the completeness limit of the SDSS imaging data. 
The comparison with deeper HST images shows that the SDSS imaging data is
50\% complete to $r^*=22.5$, which agrees with the expectation
from the simulation described above. This limiting magnitude is
substantially fainter than the $r^* = 21$
limit to which we present the galaxy counts in this paper.

For the detection of objects {\it Photo} adopts a peak finding algorithm
using the matched filter method, not only with the use of the PSF
filter but also it is applied to $2\times2$ and $4\times4$ binned image.
This algorithm is capable of detecting objects with considerably low
surface brightness. Our simulation shows that the 100\% detection 
completeness is sustained at $r^*=21$mag for galaxies with the scale 
length $5\sigma$ larger than that of $M^*$ galaxies.

\section{Construction of the Galaxy catalog}
\label{sec:galcatalog}

We have adopted two approaches to define a galaxy sample from the photometric
catalog output by {\it Photo}. For bright magnitudes 
($r^* < 16$), we have visually 
classified all objects in Runs 752 and 756 into stars and galaxies.
In \S4.3, we compare the results of this visual classification with the 
results of the star-galaxy classification algorithm used in {\it Photo}. 
At fainter magnitudes ($16 < r^* <21$), we use the star-galaxy 
classification employed by {\it Photo}. In these faint magnitudes
{\it Photo} yields 
consistent classifications of objects in different colors.

\subsection{Selection of objects from the photometric catalog}

Objects that lie in the overlap between adjacent scanlines in two strips of a
stripe, and those that lie in the overlap between adjacent frames,
appear more than once in the photometric catalog.
Each of these detections is flagged as {\tt PRIMARY} or {\tt SECONDARY}, 
based on its position in the individual frames; the detections flagged as 
{\tt PRIMARY} define a unique detection of each object. Therefore, we
construct the galaxy catalog using only the {\tt PRIMARY} objects.

{\it Photo}\  examines every detected object for multiple peaks; if they exist,
the object is deblended (Lupton et al. unpublished) into a 
hierarchy of component objects (children), 
and the parent is flagged as {\tt BLENDED}.  
If a parent object has a large number of peaks, then only the 30 highest
peaks are deblended.
If the galaxy touches the edge of a frame, or if it is larger than half the
size of a  frame, it is not deblended any further, 
and the object is additionally flagged as {\tt NODEBLEND}\footnote{Objects 
touching the edge of a frame will be deblended in the latest
versions of {\it Photo}.}. 
Only isolated objects, child objects, and objects on which the deblender
gave up are used in constructing the galaxy catalog.

\subsection{Star-galaxy classification}

{\it Photo}\ fits each object in all five colors to three model 
profiles convolved with the locally determined PSF, i.e.,
a point source, an exponential galaxy profile of arbitrary axial ratio, 
orientation and scale length,
and a de Vaucouleurs galaxy profile of arbitrary
axial ratio, orientation and scale length. For each object, 
the best-fit model in the $r'$ band is noted.
The corresponding magnitude in the different bands, 
calculated using the parameters of the best fit model in the $r'$ band
is referred to as the 
{\it model magnitude} in that band.  
In principle, it is possible to use the likelihoods of these model fits to
classify objects into stars and galaxies. We find empirically, however, 
that the star-galaxy classification is better using the following
technique.
Every object is classified as a star or a galaxy
in each band, based on the difference between the model and PSF magnitudes
\footnote{PSF magnitude is computed for all objects by fitting
a two-dimensional PSF model, which is a continuously varying function
of the position on the frame as determined from bright stars in the data.}.
An object is classified as a star in any band if the model 
magnitude and the PSF magnitude differ by no more than 0.145 mag, 
corresponding to  the fluxes in the model fit and the PSF fit to the object
satisfying the relation 
\begin{equation}
\frac{{\rm Flux\ in\ PSF}}{{\rm Flux\ in\ model}} < \frac{1}{0.875}.
\end{equation}

We have tested the accuracy of this star-galaxy classification algorithm 
by comparing our classifications of objects in SDSS images 
of the Groth strip against the classifications 
in the Medium Deep Survey catalog (MDS) constructed using WFPC2
parallel images from the {\it Hubble Space Telescope} 
\citep[and references therein]{ratnatunga99}.
The SDSS data were taken in May 2000 (Runs 1468 and 1469).
The FWHM of the median seeing
was $1.4''$, slightly worse than the FWHM in Runs 752 and 756. 
We find that our star-galaxy classification correctly reproduces
the MDS classifications for all objects  (60 stars and 68 galaxies) 
with $r^* < 20.5$. 
In the range
$20.5 < r^* < 21$, 51 out of 52 galaxies and 14 out of 14 stars are correctly
classified.
More details and tests of the star-galaxy classification algorithms used
in {\it Photo}\ can be found in Lupton et al. (unpublished).

\subsection{Bright galaxy sample}

Although the star-galaxy classification algorithm described in the previous
subsection yields correct classifications in the range 
$16 < r^* < 21$ for the purpose of deriving reliable galaxy counts, 
there are reasons to suspect that this algorithm may run into
problems at brighter magnitudes.
For example, saturated pixels and diffraction spikes can compromise the 
model-fitting algorithm\footnote{Obviously saturated pixels are rejected from
fitting, but marginallly saturated pixels may affect the model fitting.}
\footnote{We're still working on subtracting the 
diffraction spikes from saturated stars in the current version of 
{\it Photo}.}.  
In constructing the machine galaxy catalog from the photometric
catalog, we reject all objects that contain saturated pixels 
(over the entire sky, only a handful of
galaxies with active nuclei are expected to 
be saturated in the SDSS images). However, this rejection will also eliminate
real galaxies that are blended with saturated stars.
Hence, in order to calibrate the performance of the star-galaxy classification 
algorithm employed by {\it Photo}, and to understand how well the deblender
measures the magnitudes of galaxies with substantial internal substructure, 
we have visually examined the images of all objects brighter than 
$r^{*} =  16$, in the stripe defined by Runs 752 and 756, covering an
area of about 230 deg$^2$.
At these bright magnitudes and large galaxy sizes, we expect our visual 
classification to be essentially perfect.
Hence, we can use the visual galaxy catalog to quantify the completeness
and the contamination fraction of the galaxy catalog constructed by
{\it Photo}, which we refer to as the machine galaxy catalogue.
In this region, {\it Photo}\ classifies 27,304 objects 
as non-stellar (these include saturated stars), and 86,137 objects as stars.

Table 1 presents the results of comparing the visual and the machine
galaxy catalogs at $r^* < 16$.
Column (1) lists the range of  observed magnitudes (not yet corrected for 
reddening) in the $r'$ band.
Columns (2) and (3) show the number of objects in that magnitude bin 
that are classified as galaxies by visual inspection and by the machine 
galaxy catalog, respectively. 
Column (4) is the number of visually selected galaxies that are not included
in the machine galaxy catalog.
Column (5) shows the opposite case; the objects selected as galaxies
in the machine galaxy catalog that are not real galaxies upon visual
inspection.

There are 93 objects (5\% of the sample) 
that are classified as galaxies by visual inspection
but are not present in the machine galaxy catalog (the galaxies in column 4 of Table 1).
These galaxies are not included in the machine galaxy catalog for the 
following reasons:
(a) 48 galaxies are not selected because they are blended with saturated 
stars.
(b) 45 galaxies are not classified in the $r'$ band since they contain 
an interpolated pixel (i.e., a bad column or a cosmic ray)
at or near the center, a condition which causes {\it Photo} to classify 
an object as
{\tt UNKNOWN}\footnote{Objects with interpolated pixels
at the center is properly classified in later versions of {\it Photo}.}.
Among these 45 galaxies, 43 are correctly classified
as galaxies in both $g'$ and $i'$ bands. Hence, these galaxies would have been
included in the machine galaxy catalog if we had selected as galaxies those 
objects that {\it Photo}\ classifies as galaxies in at least two of the three 
bands, $g'$, $r'$, and $i'$.

Similarly, there are 215 objects that are classified as galaxies in the
machine catalog, but are not confirmed to be real galaxies upon visual
inspection (the objects in column 5 of Table 1).
These objects are included in the machine galaxy catalog for the
following reasons:
(a) 108 objects are classified as galaxies in the $r'$ band, but are
classified as stars in all the other color bands.
These objects are either single stars with slightly elongated shapes
in the $r'$ band, or spurious objects caused by satellite trails.
None of these objects are real galaxies. 
They would be excluded from the machine galaxy catalog if we required 
that the object be classified as a galaxy in at least two of the three bands
$g'$, $r'$, and $i'$.
(b) 44 objects are double stars misclassified as galaxies.
(c) 44 objects are redundant children arising from excessive deblending
of other bright galaxies already present in the visual galaxy catalog.
(d) 3 objects are groups of galaxies, whose individual members would be
fainter than $r' = 16$ if they had been properly deblended.
(e) 4 objects are selected as galaxies because of satellite trails
and classified as galaxies in at least two of the 
three colors $g'$, $r'$ and $i'$.
(f) 12 objects are ghost images due to reflections of bright stars inside
the camera.

We have re-examined the images of all the 44 spurious children which resulted 
from excessive deblending, since such deblending problems will lead to
underestimates of the flux of the parent galaxy. We have combined these 44 
objects back to 31 parents, and remeasured their Petrosian fluxes. We
find that the error in magnitude resulting from this excessive deblending
ranges from 0.1 to 1 mag, with a typical value of 0.5 mag.  We use the
corrected magnitudes for these 31 galaxies in the visual galaxy
catalog. We note that the frequency of excessively deblending objects 
decreases at fainter magnitudes; in particular, there are no such
objects in the  $15 < r^* < 16$ bin, implying that excessive deblending is
unlikely to be a problem fainter than 16 mag.

From Table 1, we see that the net statistical completeness of the machine galaxy catalog, defined as
(Galaxies common to both visual and machine catalogs)$/$(Galaxies in visual 
catalog), is $95\%$ for the $r^* \leq 16$ mag sample.
However, if we require that all objects in the machine galaxy catalog be 
classified as galaxies in at least two out of the three bands, 
the completeness increases to $97\%$.
The major reason for missing real galaxies (incompleteness) in the
machine galaxy catalog is that
they are blended with saturated stars, while most of the spurious  galaxy
detections (contaminants) are either double stars or shredded pieces
of brighter galaxies.
The galaxy number counts obtained from the machine galaxy catalog 
matches the true galaxy counts to within 3\% for the
bright sample with $r^* < 16$.

The comparison of the {\it Photo} outputs to the results of the visual
inspection has uncovered a number of subtle problems with the
photometry processing with {\it Photo} version
5.0.2 as of early 2000. Many of these problems are being fixed for a
future version of the pipeline, which should have improved performance. 

\subsection{Faint Galaxy Sample}
\label{sec:faint}

In the magnitude range $16 < r^{*} < 21$, we derive the galaxy
counts using the machine galaxy catalog.
At these magnitudes, we construct the machine galaxy catalog from the
photometric catalog using the following criteria:
(a) Only the {\tt PRIMARY} detection of an object is included in the galaxy sample.
(b) Based on the experience gained at bright magnitudes, we classify an object 
as a galaxy only if {\it Photo}\ classifies it as a galaxy in at least two of the
three photometric bands, $g'$, $r'$, and $i'$. 
(c) Objects that are blended with saturated stars are not included in the
galaxy sample.

\begin{figure}
\plotone{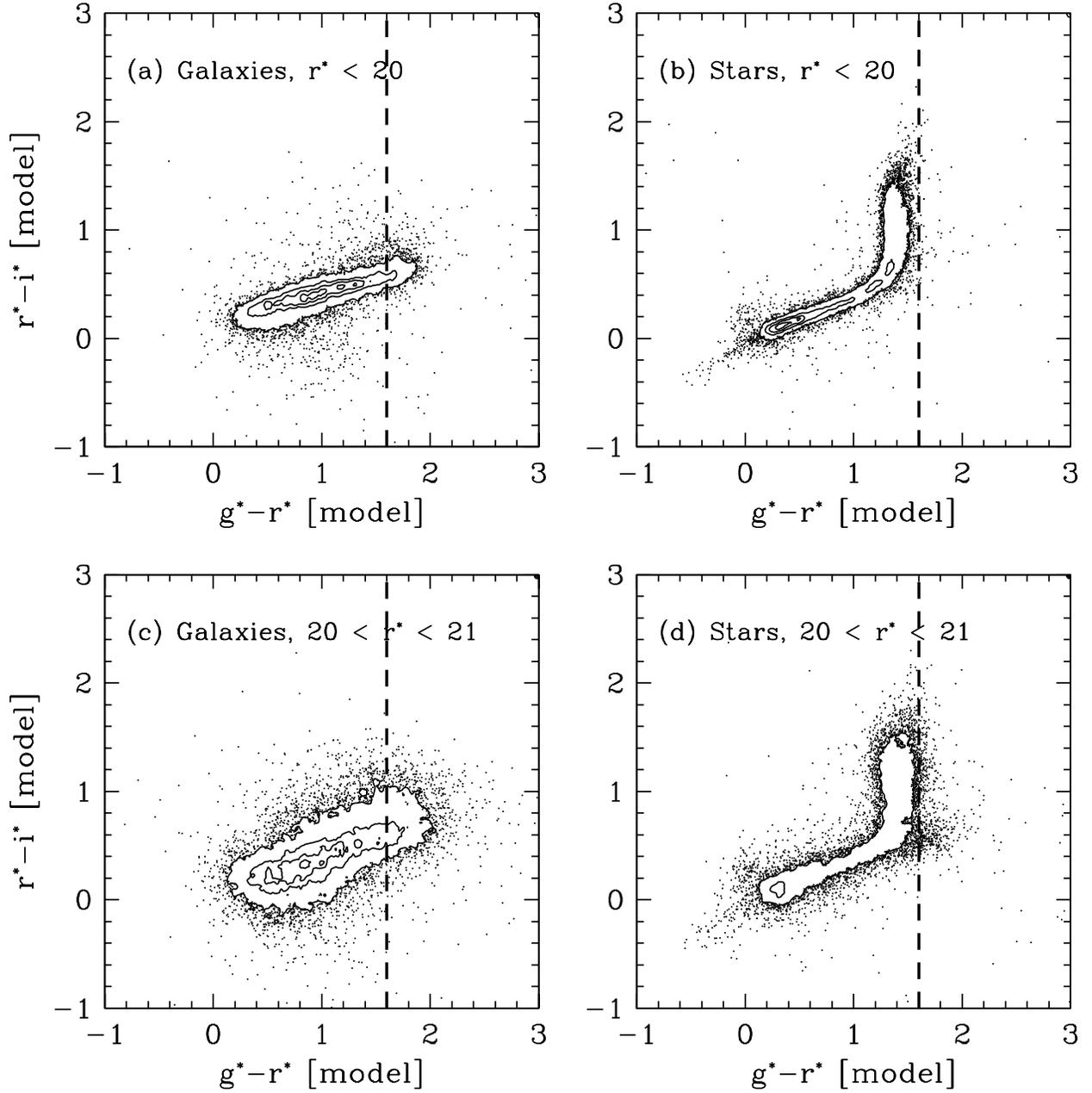}
\caption{Distribution in the $g^*-r^*$, $r^*-i^*$ plane of all objects 
classified  as galaxies and stars in fields 251 to 300 of the third CCD 
column of Run 756. (The star sample contains 50-100 quasars.) 
The inner parts of the distributions are shown as contours, linearly
spaced in the density of objects in the color-color plane.
Panels (a) and (b) show the distribution of objects classified as galaxies
and stars, respectively, in the magnitude range $16 < r^* < 20$.
Panels (c) and (d) show the corresponding objects in the magnitude range
$20 < r^* < 21$. Objects in panels (b) and (d) with $g^*-r^* > 1.6$ (shown 
by the dashed line) are real galaxies that are 
misclassified as stars, quasars, 
or stars with incorrect photometry.}
\label{color_color}
\end{figure}

Figure \ref{color_color} shows the distribution in the $g^*-r^*$ $r^*-i^*$ plane of 
all objects classified as stars and galaxies, in fields 251 to 300
of the third CCD column in Run 756 
(seeing FWHM is about $1''$)\footnote{We have not separated quasars
from stars. So we expect about 50-100 quasars being included in the
star sample. We have not studied in this paper the problem of blue
compact galaxies.}.
The colors are computed using the model magnitudes of objects in each band.
The $g^*-r^*$ color is roughly constant for stars with
$r^*-i^*\ > 0.5$ (corresponding to stars cooler than 3500 K), as the 
$g'$ and $r'$ bands are dominated by absorption from molecular bands,
especially from TiO.
Hence, the objects with $g^*-r^* > 1.6$ that are classified as stars 
(objects that lie to the right of the dashed line in panels (b) and 
(d)) are quasars,  stars with incorrect photometry, or real galaxies 
(including compact galaxies) that are 
misclassified as stars.
Even if we assume that all objects with $g^*-r^* > 1.6 (1.7)$ are really 
galaxies, only $1.4\%\  (1.0\%)$ of galaxies are being misclassified as stars 
in the magnitude range $16 < r^* < 20$, and $4\%\ (2\%)$ in the range 
$20 < r^* < 21$.
Similarly, our tests of the star-galaxy classification algorithm using the 
images from the Groth strip (described in \S4.2) show that at most a few
percent of stars are being misclassified as galaxies at $r^* < 21$.
We have also compared the classifications of objects that lie in the overlap
region between Runs 752 and 756. We found that the two classifications in 
these runs with different seeing were identical for $100\%$ of objects 
at $r^* < 20$, for $99\%$ of objects in $20 < r^* < 20.5$, and for
$97\%$ of the objects in $20.5 < r^* < 21$.
All these tests show that misclassification of objects does not pose a
serious problem in deriving the galaxy counts at all magnitudes brighter 
than $r^* = 21$.

\section{Galaxy number counts in the SDSS bands}
\label{sec:numbercounts}

Figure \ref{field_by_field_counts} shows the variation from one field to 
another of the number of 
stars and galaxies, averaged over all six CCD columns.
The different panels show this variation in different bins in $r'$ magnitude.
Figure \ref{field_by_field_counts}(a) shows the objects in Run 756, 
while Figure \ref{field_by_field_counts}(b)
shows those in Run 752.
The abscissa in each panel is the field number.
Note that a field corresponds to a $10' \times 13'$ region of sky,
and the data shown are an average over 6 fields across the stripe.
The global variation of the number of stars (on the left panels) reflects the 
change in Galactic latitude during the TDI scan over 7-8 hours. 
The top left panel in each figure shows the variation of Galactic latitude 
over the duration of the scan.
The number of galaxies varies from field to field due to both small-number
statistics and small-scale clustering. However, there are no large-scale
gradients in the galaxy distribution with field numbers in bins brighter 
than $r^* = 21$, confirming  that there is no serious contamination of stars 
to the galaxy counts in the magnitude range of interest.
The galaxy counts in the faintest magnitude bin at $r^*> 21$ could be 
systematically affected by galaxies being misclassified as stars.
In particular, note the drop in the galaxy counts for $21 < r^* < 21.5$
in Run 752 for small field numbers, where the seeing is poor,
indicating that some fraction may not be detected and more 
importantly some galaxies with low signal to noise ratio 
could be misclassified as stars.
At the magnitude where we present the galaxy counts there is no
contamination from stars to galaxies, despite
the fact that there is a higher ratio of stars to galaxies at brighter 
magnitudes.
The solid curves in the left panels show the expected variation of 
star counts from the Galaxy model of \citet{BahcallSoneira}
using the color transformations  derived from our work on primary standard 
stars \citep{smith}

\begin{eqnarray}
g^* & = & V + 0.530 (B-V) - 0.075, \\
r^* & = & V - 0.438 (B-V) + 0.113.
\label{eqngrvbmv}
\end{eqnarray}
We refer to \citet{Chen} and \citet{Finlator} for more detailed analysis 
of star counts.
These empirical color transformations differ from the 
synthetic relations derived in \citet{Fukugita:1996} by less than 0.03 mag
in the constant terms. 
The model predictions and the observed star counts agree remarkably well at 
bright magnitudes ($r^*\leq 18.5$) for all fields of the TDI
scan. For $18.5 < r^* < 20.5$, the deviations of the observed star counts from
the model predictions become increasingly apparent at large field
numbers, which are pointing toward low Galactic latitudes in the
direction of the Galactic center.

Figure \ref{r_counts} shows the galaxy number counts in 0.5 mag bins in the $r'$ band 
in the Northern equatorial stripe. We have coadded the counts in Runs 752 and 
756 for bins at $r^* < 18$. At fainter magnitudes, we present the results from
Run 756 alone. We are certainly not limited by small-number statistics at 
$r^* > 18$ mag, and therefore we do not include the galaxy counts
from Run 752, where the seeing is worse.
The solid points show the counts from the visual galaxy catalog, 
and the open points show the counts from the machine galaxy catalog.  
Star counts are also shown as crosses:
the sharp drop seen at bright magnitudes ($r^*<14.5$) is due to the fact
that we cannot measure stars that saturate in the photometric CCD.  
Thus, our star counts are reliable in the magnitude range $15 < r^* < 21$.
We deredden the Petrosian magnitudes of
galaxies using the Galactic extinction map of 
\citet{SchlegelFinkbeinerDavis}, assuming a ratio of total to
selective extinction in the 
$r'$ band $R_{r'}=2.75$\footnote{Note that Petrosian
magnitudes are easy to correct for extinction, unlike, for example, 
isophotal magnitudes (see the discussion in \citet{Santiago96}).}
computed from a standard extinction law \citep{SchlegelFinkbeinerDavis}. 
For most of the sky region we have scanned, extinction is smaller
than 0.1 mag (Figure \ref{extinction} shows selective extinction 
$E(B-V)$  
along the Celestial Equator).
A smooth match between the galaxy counts from the visual and the machine
galaxy catalogs is apparent; the two samples give almost identical galaxy
counts in the range of magnitudes $14 < r^* < 15.5$.\footnote{The 
visually selected galaxy catalog is no longer complete
in the magnitude bin $15.5 < r^* < 16 $ after reddening corrections.}
Hence, we derive the galaxy counts from the visual galaxy catalog at 
$r^* \le 15.5$, and from the machine galaxy catalog at fainter 
magnitudes.
Table 2 presents the galaxy counts in the five SDSS bands 
$u',\ g',\ r',\ i',$ and
$z'$. We list the actual number of galaxies ${\cal N}$ detected in our 
sample in each 0.5 mag bin (after dereddening the magnitudes of galaxies),
together with the effective area in which we searched for galaxies in that
bin.
We note that the effective areas in the magnitude bins from the machine galaxy 
catalog are corrected for regions which are masked to avoid bright stars 
(amounting to about 0.4\% of the total area). 

The error bars on the galaxy counts data in Figure \ref{r_counts} include
contributions from both Poisson noise and large scale structure in the
galaxy distribution.
The covariance between the galaxy counts in two  magnitude bins 
is \citep{peebles93}
\begin{equation}
\left< \delta N_{i}\delta N_{j} \right> = N_{i}\delta^K_{ij} +
N_{i}N_{j}\int\!\int \frac{w_{ij}(\theta_{ij})d\omega_{i}d\omega_{j}}{\omega_{i} \omega_{j}},
\label{eqn:5}
\end{equation}
where $N_i, N_j$ are the number of galaxies in bins $i$ and $j$,
$\omega_{i},\ \omega_{j}$ are the areas of the sky from which the galaxy
numbers are derived, $\delta^K_{ij}$ is the Kronecker delta symbol,
$w_{ij}(\theta)$ is the angular correlation function between the 
galaxies in bins $i$ and $j$, and the double integral extends over every
pair of points in areas $\omega_{i}$ and $\omega_{j}$ that are separated
by any angle $\theta_{ij}$.
The first term is the contribution from Poisson noise and is included only in 
computing the variance in counts in a magnitude bin.
The second term is the contribution from large-scale structure in the 
galaxy distribution. We use the covariance matrix of the galaxy counts
in different magntiude bins in fitting for the normalization of
the counts-magnitude relation.

In any two magnitude bins, we estimate the contribution from the
second term as follows.
We first assume that $w(\theta)$ can be represented as a power-law 
of the form $w(\theta) = A(m)\theta^{1-\gamma}$, where $\gamma=1.8$ is
the slope of the two point correlation function \citep{peebles93}.
We describe the details of computing the angular correlation
function $w_{ij}(\theta)$ between galaxies in different magnitude bins
in Appendix A. 
We perform the double integral 
\begin{equation}
I_{ij} = \int\!\int \theta_{ij}^{1-\gamma}d\omega_{i}d\omega_{j}
\end{equation}
in a patch of sky $90^\circ \times 2.5^\circ$, roughly corresponding
to the size and geometry of the stripe defined by Runs 752 and
756, and use the same value of $I_{12}$ for all possible pairs
of geometries of the survey areas in different magnitude bins. 
We fix the amplitude of $w(\theta)$ assuming that 
$w(\theta = 0.1^\circ) = 0.1$ for galaxies in the magnitude range
$18 < r^* < 18.5$, approximately the value measured from SDSS
commissioning data itself \citep{scranton}.

The line segment in Figure \ref{r_counts} shows a fit to the galaxy 
counts-magnitude relation expected in a homogeneous universe 
assuming Euclidean geometry for 3 dimensional space (we refer this as 
``Euclidean'' geometry in the following text),
\begin{equation}
N(m_{\lambda})=A_{\lambda} 10^{0.6(m_{\lambda}-16)}.
\label{eqn:7}
\end{equation}
The observed galaxy counts are quite consistent with this line 
with $A_{r^*}=5.99\pm 0.52$ (0.5mag)$^{-1}$deg$^{-2}$
($\chi^{2} = 13.9$ for 9 degrees of freedom)
for $12 < r^* < 17$ and even beyond this range. 
For $r^* <  13$, the number of galaxies is 
quite small ($N\leq 20$ in each bin), and there are large errors due
to both Poisson noise and large-scale clustering.

In order to study the number counts in detail, we plot in 
Figure \ref{r_growth} the quantity $N(r^*)10^{-0.6(r^*-16)}$, the observed
galaxy number counts normalized by the shape of the counts-magnitude
relation (hereafter, called {\it the growth rate})
expected for a homogeneous galaxy distribution in a universe with
``Euclidean'' geometry.
We also show the predictions of a no-evolution model, taking
into  account both cosmological corrections and K-corrections.  
The model is based on studies of the statistics of galaxy parameters
in the $B$ band. We use a mix of E, S0, Sab, and Scd galaxies, in the
ratio of morphological types given in \citet{Fukugita:1998}
and use the color shifts and K-corrections given in \citet{Fukugita:1995}.
We assume that the luminosity function is characterized by a Schechter
function with parameters $\alpha=-1.1$, and $M_B^*=-19.5+5\log h$.
The shape of the predicted curves depends only weakly on the model 
parameters. 
We show the predictions of the no-evolution model in three different 
cosmologies: 
$(\Omega_{m}, \Omega_\Lambda)=(1, 0),\ (0.3, 0)$, and $(0.3, 0.7)$.
We also show the predictions in a universe assuming Euclidean geometry
for three space, but including K-corrections.
Model curves are normalized to the amplitude $\tilde A_{r^*}$ derived in 
equation~(\ref{eqn:8}) below.

The observed galaxy counts are consistent with the predictions of the
no-evolution model for $r^* > 16$, although the data permit some 
evolution.
At $r^*<16$, the data mostly lie below the predictions of the 
no-evolution model, but they are mostly 
within the fluctuations expected from
Poisson noise and large-scale structure in the galaxy distribution.
We will show in \S~\ref{sec:counts_otherbands} that this deficit of
galaxies at bright magnitudes is much smaller in amplitude and much
more local in volume
than the local underdensity claimed by the APM group \citep{Maddox}. 
This small underdensity toward the Northern Galactic cap is also not 
seen in our data toward the Southern Galactic cap (analyzed in 
\S~\ref{sec:south} below). It is  therefore likely to be ascribed to
a local 
fluctuation due to large scale structure in the galaxy distribution.

We repeat the same analysis of the galaxy counts in the $u'$, $g'$, $i'$ and
$z'$ passbands.
We assume the ratio of total to selective extinction to be 
$R_u'=5.16$, $R_g'=3.79$, $R_i'=2.09$ and $R_z'=1.48$, respectively, 
taken from a standard extinction law \citep{SchlegelFinkbeinerDavis}.
The results are presented in Figure \ref{all_counts} and Table 2.
The error bars include contributions from both Poisson noise and large
scale structure. In each band we estimate $w(\theta)$ of galaxies
in a magnitude bin from the corresponding value of $w(\theta)$ in the 
$r'$ band, assuming a mean color for galaxies in that magnitude bin.
The galaxy counts in the $g'$, $i'$ and
$z'$ passbands
closely follow the $10^{0.6m}$ law at bright
magnitudes $(m < 17-18)$, as we have seen with the counts in the
$r^*$ band.
The counts in the $u'$ passband show some
wiggle for $u^*<18$ mag, and the local deficit is more
conspicuous for brighter magnitudes. The reason is not clear to us. 

As is evident from Figure \ref{r_growth}, cosmological corrections and 
K-corrections have a non-negligible effect on the galaxy counts
at magnitudes even brighter than $r'=16$ mag (as seen from the 
difference in shape between 
the curves corresponding to different cosmologies and the ``Euclidean''
geometry), although in the data, this effect is masked by fluctuations 
arising from
Poisson noise and large-scale structure.
To obtain the cosmologically meaningful
normalization of the galaxy counts, we must take account of the
cosmological and K-corrections. 
The fit must cover a magnitude range faint enough that local large
scale structure effects are avoided, but not so faint that 
evolutionary effects become important.
In Figure \ref{Afitting} (c) we display with solid points 
the galaxy count data in the
$r'$ band divided by the no-evolution model prediction
(i.e. the ``Euclidean'' growth factor and the cosmological and K-corrections)
so that data should fall on a constant line if they are not affected
by large scale structure and evolution. Here we assume
$(\Omega_{m}, \Omega_\Lambda)=(0.3, 0.7)$, although the results are
insensitive to cosmology (The open points are the counts data from
Southern Celestial Equator stripe, which are discussed in \S\ref{sec:south}).
We fit the data (only those denoted by solid points)
taking account of Poisson noise and large-scale structure
using the fitting range chosen to cover the plateau.
We have carried out both full correlated and uncorrelated fits.  The
result of full correlated fit, however, turns out to give a rather
poor fit to the number count data\footnote{This implies that the smooth
no-evolution model used here is not quite suitable to describe the
data (Correlated fits are sensitive to the adequacy of the fitting
function).  The values of $\tilde A$ from the full correlated fit are
systematically lower by $0.6-1.3\sigma$ (except for $u'$) than the
values from the uncorrelated fit, which we have given in the figure
and table. The fitting function for $u'$ is apparently inadequate, and
we find a larger discrepancy between the two fits. For completeness,
the values of $\tilde A$ from full correlated fit for $u^*$, $g^*$,
$r^*$, $i^*$, $z^*$, and $B$ band are $0.91\pm0.10$, $4.73\pm0.24$,
$10.47\pm0.57$, $17.48\pm0.81$, $28.71\pm1.17$, and $2.78\pm0.19$,
respectively.}. Hence, we adopt the result from the uncorrelated fit.
For example for the $r'$ band,
we obtain the coefficient of the ``Euclidean'' growth rate
in the bright limit, which we denote by $\tilde A_{r^*}$, 
\begin{equation}
\tilde A_{r^*}= 11.30\pm 0.75 \,(0.5~{\rm mag})^{-1} \,\rm deg^{-2}\ .
\label{eqn:8}
\end{equation}
where the error includes Poisson noise, large scale structure and
the variation induced by the change of the faint end of the fitting
range by $\pm 1$ mag.
Figure \ref{Afitting} presents 
the analysis repeated for all passbands (including $B$ band which we discuss
in the next section). 
The $1\sigma$ error band is indicated with dotted lines in the figure.
We obtain the coefficient of the ``Euclidean'' growth rate
in the bright limit $\tilde A$ in each band, which is
given in Table \ref{table3}, 
together with the range of magnitudes over
which it was fit. We could not find a plateau for the $z'$ band
(to a lesser extent for the $i'$ band), 
and hence the
real error for $\tilde A(z^*)$ would be larger than is
quoted in the table.

We do not discuss the detail of star counts in this paper.  
Extensive analyses of the star counts in the SDSS bands, and their
implications for models of Galactic structure are presented in
\citet{Chen} and \citet{Finlator}.

\section{Galaxy number counts in the $B$ and $I_{814}$ passbands}
\label{sec:counts_otherbands}

The majority of results on galaxy number counts that exist in the literature
have used the $B$ (or $B_J$) band. Therefore, in this section, we
convert our galaxy counts results to the Johnson-Morgan $B$ band
photometric system, 
and determine the normalization
of the counts-magnitude relation.
We adopt the color transformation
\begin{equation}
B-g^* = 0.482 (g^*-r^*) + 0.169,
\label{eqn:9}
\end{equation}
derived from preliminary standard star work \citep{smith}
for the Landolt stars \citep{Landolt} with the standard deviation of 0.03 mag.
We find that equation~(\ref{eqn:9}) provides a better match to the
observed $B$ band photometry than an equivalent expression using the 
$u^*-g^*$ color\footnote{One may consider the transformation to $B$ magnitude 
via $u^*$ and $g^*$, since the $B$ band is located between 
the $u^*$ and $g^*$ bands. However, this
gives a larger offset except for late type spiral galaxies;
for example, for $g'-r'\simeq 0.7$ the offset is 0.06 mag.
This offset results from the difference of $u'$ band spectral features of
stars and galaxies.}.  
We have checked the accuracy of this transformation using colors
synthesized from the spectra of galaxies.
We use the spectroscopic energy distributions of galaxies from the 
spectrophotometric atlas of \citet{Kennicutt}, and carry out synthetic 
calculations of $u',\ g',\ r'$ and $B$ magnitudes (see \citet{Fukugita:1995}
for more details).
The $B$ magnitude calculated via equation~(\ref{eqn:9}) from synthetic 
$g'$ and $r'$ magnitudes 
agrees with the synthetic $B$ with an offset of $-$0.01 to +0.04 mag
for elliptical and Sb galaxies at zero redshift (+ sign means
that the color transformation makes $B$ magnitude brighter than the
synthetic value). 
The offset increases to +0.1 mag for late type spirals. As we
will see in \S~\ref{sec:colors} below, the mean color of our bright
galaxy sample is $g^*-r^*\sim 0.7$, for which the offset nearly vanishes.
The color transformation will in general also 
depend on the redshift of the galaxy, due to K-corrections.
The accuracy degrades when the 4000\AA~spectral break moves deep
into the $g'$ band, but we find that the color transformation
given by equation~(\ref{eqn:9}) works as well at
$z \approx 0.2$ as at $z \approx 0$; for a galaxy at 
$z\approx 0.4$ the offset increases
to about $0.2$ mag.  
We apply the color transformation in equation~(\ref{eqn:9})
on a galaxy by galaxy basis, and show the results only for $B < 19.5$,
as the fainter magnitude bins include a substantial fraction of
galaxies with $z>0.3$. 

Figure \ref{B_counts} shows the galaxy counts in the transformed $B$ band 
(after correcting for Galactic reddening), and Figure \ref{B_growth} shows the 
same quantity after normalizing by the ``Euclidean'' growth factor of
$10^{0.6m}$.
The latter figure also shows the galaxy counts data in the $B$ band from
the APM survey \citep{Maddox}, and from the surveys of 
\citet{Gardner} and \citet{BertinDennefeld}, together with the 
predictions of the no-evolution model. The photographic $b_J$ band data of
\citet{Maddox} and \citet{BertinDennefeld} have been converted to
$B$ band by \citet{Gardner}.  Note that each of these studies
uses a different method to determine the magnitudes of galaxies, although
there are attempts to make corrections to obtain pseudo-total
magnitudes;  we refer the reader to the respective papers for details.

The shape of our galaxy counts fainter than $B = 16.5$ agrees well with
those of  \citet{Gardner} and \citet{BertinDennefeld}, but
our normalization is about 15\% higher than theirs.
On the other hand, the APM data lie substantially below our counts
brighter than 18.5 mag.  
In particular, in the magnitude range $15 < B < 18$, our galaxy counts
are on average a factor of two larger than the counts inferred from
the APM survey.
The shape of our counts-magnitude relation is also consistent with the
predictions of the no-evolution model in the magnitude range $17 < B <
19$, while some deficit is seen in magnitudes brighter than $B=16$
arising from local large scale structure in the galaxy distribution. 
Since the depth corresponding to the deficit is only 
$150 h^{-1}$Mpc$^{-1}$ and our survey is only in a small patch on the
sky, it appears that the deficit brighter than $B=16$ is consistent with
the amplitude of large-scale structure fluctuations. We need not invoke
an overall local underdensity of galaxies over a large solid angle.
This is corroborated by the analysis for the Southern stripe, as we
discuss later. 
We note that 
the APM data remain flat over the range from 15.5 to 19 mag, while
cosmological effects and K-correction should make the counts 
decrease significantly over this magnitude range; Maddox et
al. (1990) interpret this as due to evolution.
Our data do not show evidence for such a rapid `evolution' of the galaxy
population to $z = 0.2$ (approximately the redshift of an $L^*$ galaxy 
at $B=19$). 

The star counts plotted in Figure \ref{B_counts} agree well with 
the predictions of the Bahcall-Soneira model in the range $15 < B < 19$. 
The agreement is a qualitative confirmation that our color
transformation in equation~(\ref{eqn:9}) works well for stars.

We have also calculated the galaxy number counts in the $I_{814}$ passband 
(F814W) --- the band in which much work has been done with the 
{\it Hubble Space Telescope}.
From our standard star work \citep{smith}, we find 
for the transformation to the Cousins' $I_c$ band that
\begin{equation}
I_c-i^*=-0.205 (r^*-i^*)-0.382,
\label{eqn:10}
\end{equation}
with the standard deviation of 0.02 mag.
We further apply color transformation from $I_c$ to $I_{814}$:
$I_{814}=I_C+0.04$ according to \citet{Fukugita:1995}. This constant
depends little on galaxy colors.

Figure \ref{I_counts} presents our galaxy counts in the $I_{814}$ band for
$I_{814} < 20$ mag.  
We also plot the data from \citet{Gardner} and \citet{Postman}.
The three independent data sets show good agreement up
to an offset of about $0.05$ mag, which might be ascribed to systematic errors
in the different bands or to different definitions of the magnitudes
of objects.  Our numerical data for the galaxy counts are given in 
Table \ref{table4} for both $B$ and $I_{814}$ bands.

\section{Local luminosity density of the universe}

The normalization of the number counts 
$A$ of eq.~(\ref{eqn:7}) is related to the
parameters of the Schechter luminosity function \citep{Schechter} as
\begin{equation}
A=0.7046{\omega\over 3}(d^*_{16})^3\Gamma(\alpha+5/2)\phi^*,
\label{eqn:11}
\end{equation}
where $d^*=10 {\rm pc} \times 10^{0.2(16-M^*)}$, and $\omega$ is the
solid-angle coverage of the sample. 
When we take account of cosmological effects 
and K-corrections, $A$ must be
replaced with the coefficient $\tilde A$ obtained 
after corrections as we saw in \S\ref{sec:numbercounts}.
We then compute $\phi^*$ using equation (\ref{eqn:11}). 

Let us calculate the luminosity density of the universe in the
$B$ band, which has long suffered from significant uncertainties.
Since we do not determine $M^*$ and $\alpha$ in the $B$ band 
from our own data, we
use the measurements from the literature
\citep{Efstathiou,Loveday,Zucca,Ratcliffe,Folkes}.
These parameters 
lie in the range $-19.7\leq M_B^*\leq -19.4$ and
$-1.0\geq\alpha\geq-1.2$ \footnote{\citet{Folkes} give $\alpha$
significantly steeper than this range. Their Schechter fit, however, poorly
represents the step-wise maximally likelihood result, and the faint end
slope is significantly flatter than their $\alpha$ indicates.},
where $B_J=B+0.1$ is applied to convert into the $B$ band when necessary.
From $\tilde A_B=2.98\pm0.17$ ($B=12-19.5$), we obtain 
$\phi^*=2.05\pm0.12{+0.66\atop -0.28}$. Here the first error
stands for that of $\tilde A$ and the second arises from
uncertainties in the Schechter parameters.

We then calculate the luminosity density of the universe in the $B$ band.
Note that the fractional error in luminosity density
(${\cal L}$) is much smaller than that in
$\phi^*$, because the luminosity density goes as 
${\cal L} \sim \phi^*L^*\Gamma(\alpha+2)$
while the number counts go roughly as 
$N(m)\sim \phi^*L^{*3/2}\Gamma(\alpha+5/2)$, so that errors in 
$L^*$, $\alpha$  and $\phi^*$ largely cancel. We find that the
error in ${\cal L}$ produced by uncertainties in the 
Schechter parameters is reduced to 3\%.
The error in ${\cal L}$ is dominated by the error in the 
global number counts. We obtain 

\begin{equation}
{\cal L}_B=(2.41\pm0.39)\times10^8 L_\odot h ({\rm Mpc})^{-3}.
\end{equation}
as our best estimate. 
Here we have applied a 3\% upward shift for the expected offset
between our Petrosian flux and the total flux
(see the discussion in \S~\ref{sec:photometry}), and included a
5\% error  from  provisional photometric calibration and uncertainties 
in the color transformation equations.  

We also calculate the luminosity density of the universe in the
five SDSS passbands 
$u',\ g',\ r',\ i'$ and $z'$, in a similar manner,
adopting the values of the 
Schechter luminosity function parameters derived from the SDSS commissioning
data itself (Blanton et al. 2000).
Table 5 presents our estimates of $\phi_{\lambda}^*$ and ${\cal L}_{\lambda}$
in each band (in units of the solar luminosity in the respective band).
We adopt $M_{u'}(\odot)=6.38$
$M_{g'}(\odot)=5.06$, $M_{r'}(\odot)=4.64$, $M_{i'}(\odot)=4.53$
and $M_{z'}(\odot)=4.52$ which roughly correspond to $(B-V)_\odot=0.63$
(Fukugita et al. unpublished). Here we assume the covariance of
the errors of the Schechter fit, 
$\Delta (M_B \cdot \alpha)/\sqrt{(\Delta M_B)^2 \cdot (\Delta \alpha)^2} \approx 1$.
The values of $\phi_{\lambda}^*$ and ${\cal L}_{\lambda}$ are consistent
with those of \citet{Blanton} within our $1\sigma$ error for $r', i' and z'$.
Our values are higher for $u' and g'$, but within $2\sigma$.

\section{Galaxy number counts in the Southern equatorial stripe}
\label{sec:south}

We have also analyzed the galaxy counts in an additional equatorial
stripe  toward the Southern Galactic cap.  These data were taken on 
September 19 and 25, 1999 (SDSS Runs 94 and 125). The first run
scanned the sky for 5.3 hours from  $\alpha =336.7^\circ$ to
$56.28^\circ$, with declination extending from $-0.^\circ 93$ to $+1.^\circ17$.
The second run scanned for 5.7 hours from $\alpha =350.^\circ6$ to
$76.^\circ7$ with declination varying from $-1.^\circ14$ to $+0.^\circ96$;
the two runs together make a filled stripe.  
The median seeing FWHM in the $r'$ band was $1.4''$ in Run 94 and $1.7''$ in 
Run 125, but the telescope was not well collimated during these runs.
The total area covered in this Southern equatorial stripe is 210 deg$^2$. 
We do not construct the visual galaxy catalog in this region, 
but instead use the machine galaxy catalog itself over the entire
range of magnitudes $11 < r^* < 21$.
For $r^* > 18$, we present the galaxy counts from Run
94 alone, since it has data taken under better seeing.
We present the galaxy counts in the $u'$, $g'$, $r'$, $i'$ and $z'$ bands in 
Figure \ref{all_counts_south} and Table 6.

Figure \ref{r_growth_with_south} shows the galaxy number counts in $r'$ band normalized by 
the ``Euclidean'' growth factor of $10^{0.6m}$, in both the Northern
(solid points) and the Southern (open points) equatorial stripes.
The error bars on the galaxy counts in both stripes include
contributions from both Poisson noise and large scale structure.
The number counts from these two essentially independent regions of
the sky are identical for $r^*\geq 16$ mag. At brighter magnitudes,
the counts in the Southern stripe are systematically larger than those
in the Northern stripe. The Southern stripe crosses the southern
extension of the Pisces-Perseus supercluster; the enhanced counts are
due to this well-known overdensity.  
The curves represent the predictions of the no-evolution model 
we described earlier.

The degree of disagreement in the galaxy counts at $r^* < 16$ between 
the Northern and the Southern  stripes indicates that the
deficit of galaxies seen in the Northern stripe is a local effect
that depends on direction in the sky; i.e., the local universe to 100$-$200
Mpc is clumpy. In the magnitude range $14 < r^* < 15.5$, there are
$N_{1} = 1010$ galaxies in the Northern stripe, in an area of 
$\omega_{1} = 230$ deg$^{2}$.
The rms fluctuation in this number due to large scale structure is
computed using equation~(\ref{eqn:5}) to be $\sigma  =
170$. In the same magnitude range, there are $N_{2}=1259$ galaxies in
the Southern stripe, in an area of $\omega_{2} = 210$ deg$^{2}$. 
The difference in counts between the two stripes is therefore
$|N_{1}-N_{2}\omega_{1}/\omega_{2}|= 369$ --- plausibly a $2\sigma$ fluctuation
due to large scale structure in the galaxy
distribution. Thus, the difference in bright galaxy counts
between the Northern and Southern equatorial stripes can be 
explained as arising from a large scale structure
fluctuation in the galaxy distribution.
This supports our earlier observation in \S~\ref{sec:counts_otherbands}
that the deficit of galaxies at 
bright magnitudes in the Northern stripe is due to large scale
structure fluctuations.
We have to wait, however, for the end of the survey to draw
definitive and quantitative conclusions on this point, as we
quantify large scale structure over a much larger sample.

We refer to Figure \ref{Afitting} above for a comparison of
the counts in the Northern and Southern equatorial stripes for 
the other color bands.

\section{Color distributions of galaxies}
\label{sec:colors}

Figures \ref{hist_g_r} and \ref{hist_r_i} show the distribution of $g^*-r^*$ and $r^*-i^*$
colors of all galaxies in the third CCD column of Run 756 as a representative
sample.
Dereddened Petrosian magnitudes are used to calculate colors,
and color distributions are shown according to magnitude bin
in $r^*$. Note that the Petrosian magnitude in each color band is computed
using the Petrosian aperture calculated in the $r'$ band.
Therefore, the colors are measured through a consistent aperture in
all bands.
The mean ($\mu$) and rms scatter ($\sigma$) are indicated
in each panel. 

The mean of $g^*-r^*$ colors for the brightest bin is 0.71,
which is slightly (0.04 mag) redder than expected from a spectroscopic
synthesis calculation for a standard morphology mix (Fukugita et al. 1995). 
Figure \ref{hist_g_r} shows that the mean $g^*-r^*$ colors of galaxies become
systematically bluer at bright magnitudes; the mean value of
$g^*-r^*$ gradually shifts to 1.03 at
$20 < r^* < 21$.
The mean $r^*-i^*$ color of galaxies (0.36 in Figure \ref{hist_r_i})
is consistent with what is expected from
a synthesis calculation. This color  
stays roughly constant over the entire range of magnitudes
$15 < r^* < 21$, as expected from the 
spectroscopic energy distributions of normal galaxies. 

The dispersions in both the color distributions increase towards
faint magnitudes. The scatter in color in bin $20 < r^* < 21$ is
0.52 mag in $g^*-r^*$, and 0.35 mag in $r^*-i^*$.
These scatters are much larger than those expected from photometric
errors alone; the photometric error in colors in this magnitude range
is about 0.10 mag in $g^*-r^*$, and 0.07 mag in $r^*-i^*$, both as
quoted by {\it Photo}\, and from tests using simulated images.
This broadening of the color distributions is expected from the
K-corrections, which become increasingly different between galaxies of
different morphological types with increasing redshift, and also from
the broader range of redshifts that contribute to
the colors in a given magnitude bin at fainter
magnitudes.

\section{Conclusions}
\label{sec:conclusions}

We have presented the number counts of galaxies in five SDSS bands,
$u'$, $g'$, $r'$, $i'$ and $z'$, in the magnitude range $12 < r^* < 21$.
We have first examined in detail the photometric catalog derived from automated
software, {\it Photo}.
We have carefully analyzed the bright galaxy sample by comparing the galaxy
catalog produced by {\it Photo} against that produced by eye inspection.
We have shown that the statistical completeness of the machine generated
galaxy catalog is 97\% for the sample with $r^*<16$ mag. 
However, 5\% of objects in the machine galaxy catalog up to $r^*=16$ mag 
are not real galaxies.
Toward the faint end, the imaging data are complete to $r^*=22.5$ mag, and 
our star-galaxy classification is 100\% accurate to $r^*=20$ mag. 
The misclassification rate is still a few percent at $r^*=21$ mag. 
We expect that the performance will be improved in the future version
of {\it Photo}.

We have then presented the number counts of galaxies in five SDSS bands,
$u'$, $g'$, $r'$, $i'$ and $z'$, in the magnitude range $12 < r^* < 21$.
The galaxy number 
counts are derived from two independent stripes of imaging data 
along the Celestial Equator, one each toward the North and the South 
Galactic cap, each covering about 200 deg$^2$ of sky. The galaxy counts
from both stripes agree very well in the magnitude range $ 16 < r^* < 21$,
implying that we are sampling galaxies in a volume that is large enough to
be a  fair sample of the universe at these magnitudes.
At brighter magnitudes $(14 < r^* < 16)$, the galaxy counts from the two 
stripes differ by about $30\%$.  This difference is consistent
(at the $2\sigma$ level) with the fluctuations arising from large scale 
structure in the galaxy distribution.
Additionally, we have used empirically determined color transformations 
to translate galaxy counts in the SDSS color bands into those in 
the $B$ and $I_{814}$ bands. We obtained the $B$ band local luminosity
density, ${\cal L}_{B} = 2.4 \pm 0.4 \times 10^8L_\odot h $Mpc$^{-3}$,
for a reasonably wide range of parameters of the Schechter luminosity 
function in the $B$ band.

At the bright magnitudes where cosmological and evolutionary
corrections are relatively small, the shape of the galaxy number 
counts-magnitude relation is well characterized by 
$N(m_{\lambda})\propto 10^{0.6m_{\lambda}}$, the relation 
expected for a 
homogeneous galaxy distribution in a universe with ``Euclidean''
geometry. We have used the amplitude of this relation, $\tilde A_{\lambda}$, to
determine the luminosity density of the universe at zero
redshift in all the five SDSS bands and in the $B$ band
(see Table 5).

The results presented in this paper are based on
only a fraction of the imaging data taken during the commissioning
phase of the SDSS. However, given the high rate of imaging of the SDSS
(of about 20 deg$^{2}$ per hour), we already have large sufficient sky
coverage from only a few nights of observations to place interesting
constraints on such fundamental quantities as the counts-magnitude
relation and the luminosity density of the universe.
The uniform photometric accuracy and large dynamic range of the CCDs,
coupled with the high rate of imaging of the SDSS will soon enable
studies of both the properties and the large scale distribution of
galaxies at unprecedented accuracy.

The errors in the galaxy number counts and the luminosity density of
the universe in different bands derived in this work are still
dominated by fluctuations due to large 
scale structure.  
The full SDSS survey will increase the sky coverage by a factor of 25, and more
importantly enable the study of field to field variation of bright
galaxy counts. Since the fractional error in galaxy counts due to large scale
structure is inversely proportional to sky coverage, we can
expect to reduce this error (and the error in the luminosity density
of the  universe) by a similar factor.

\acknowledgments

MF acknowledges support in Tokyo from a Grant in Aid of the
Ministry of Education of Japan, and in Princeton from the Raymond and Beverly
Sackler Fellowship.
MAS acknowledges support from Research Corporation and NSF grants
AST96-18503 and AST-0071091.
IS was supported by the NASA grant NAG5-3364.
The Sloan Digital Sky Survey (SDSS) is a joint project of The
University of Chicago, Fermilab, the Institute for Advanced Study, the
Japan Participation Group, The Johns Hopkins University, the
Max-Planck-Institute for Astronomy, New Mexico State University,
Princeton University, the United States Naval Observatory, and the
University of Washington.  Apache Point Observatory, site of the SDSS
telescopes, is operated by the Astrophysical Research Consortium
(ARC).
Funding for the project has been provided by the Alfred P. Sloan
Foundation, the SDSS member institutions, the National Aeronautics and
Space Administration, the National Science Foundation, the
U.S. Department of Energy, Monbusho, and the Max Planck Society.  The
SDSS Web site is http://www.sdss.org/.

\appendix
\section{Angular cross-correlation between galaxies in different magnitude bins}

The relation between the angular two-point auto-correlation function
$w_{a}(\theta)$  computed from a flux-limited galaxy photometric catalog,
and the spatial two-point correlation function $\xi(r)$,
is \citep{limber53,peebles93}
\be
w_{a}(f,\theta) = \frac{\int \int
(r_{1}r_{2})^{2}dr_1dr_2\psi(r_1)\psi(r_2)\xi(r_{12})}
{[\int r^{2}dr\psi(r)]^2},
\label{eqn:wint}
\ee
where $f$ is the flux limit of the photometric catalog,
$r_{12} = \sqrt{r_1^2 + r_2^2 + 2r_1r_2\cos(\theta)}$ is the
distance between the points $r_{1}$ and $r_{2}$, and
the selection function $\psi(r)$ is related to the galaxy
luminosity function $\phi(L)$ by
\be
\psi(r) = \int_{4\pi r^{2}f}^{\infty}\phi(L)dL.
\label{eqn:psi}
\ee
Note that in deriving equation~(\ref{eqn:psi}) we have assumed that 
the galaxy clustering is independent of luminosity.
The angular auto-correlation function $w_{a}(f_1,f_2,\theta)$ of
galaxies with fluxes in the range $f_2 < f < f_1$ is similar to
equation~(\ref{eqn:wint}), but with the selection function defined by
\be
\psi(r) = \int_{4\pi r^{2}f_2}^{4\pi r^{2}f_1}\phi(L)dL.
\ee

The two-point angular cross-correlation between galaxies in two
different magnitude bins corresponding to flux ranges 
$f_2 < f < f_1$ and $f_4 < f < f_3$ is
\be
w_{c}(f_{1},f_{2},f_3,f_4,\theta) = \frac{\int \int (r_{1}r_{2})^{2}dr_1dr_2\psi_1(r_1)\psi_2(r_2)\xi(r_{12})}{\int r^{2}dr\psi_1 \int r^{2}dr\psi_2 }.
\label{eqn:wcross}
\ee
The two selection functions are
\be
\psi_1(r) = \int_{4\pi r^{2}f_2}^{4\pi r^{2}f_1}\phi(L)dL, \ \ 
\psi_2(r) = \int_{4\pi r^{2}f_4}^{4\pi r^{2}f_3}\phi(L)dL.
\ee
For equally spaced magnitude bins, $(f_4/f_2) = (f_3/f_1)$, 
\be
\psi_{2}(r) = \psi_1(r\sqrt{f_4/f_2}).
\ee
For a power-law correlation function of the form 
$\xi(r_{12}) = (\frac{r_0}{r_{12}})^{\gamma}$,
equation~(\ref{eqn:wcross}) becomes
\be
w_{c}(f_{1},f_{2},f_3,f_4,\theta) = \frac{r_{0}^{\gamma} \int \int
(r_{1}r_{2})^{2}dr_1dr_2\psi_1(r_1)\psi_1(r_2\sqrt{f_4/f_2})
r_{12}^{-\gamma}}{\int r^{2}dr\psi_1(r) \int
r^{2}dr\psi_1(r\sqrt{f_4/f_2})}.
\label{eqn:wcross2}
\ee
Under the small-angle approximation ($\vert \theta \vert \ll 1$), 
\be
r_{12} = \sqrt{ ( r_2- r_1 )^2 + r_1r_2\theta^{2} }.
\ee
The numerator in equation~(\ref{eqn:wcross2}) is dominated
by small values of $r_{12}$. 
So, defining 
\be
\nu = \frac{r_2-r_1}{r_0},
\ee
and setting $r_1 = r_2 \equiv r$ in the numerator of 
equation~(\ref{eqn:wcross2}), we get 
\be
w_{c}(f_1,f_2,f_3,f_4,\theta) \approx 
r_0^{\gamma} \left( \frac{f_4}{f_2} \right)^{3/2}
\left[ \int r^2dr\psi_1(r) \right]^{-2}
\int r^{4} \psi_1(r)\psi_1(r\sqrt{f_4/f_2})dr
\int \frac{d\nu}{\left[ \nu^2 + r^2\theta^2 \right]^{\gamma/2}} 
\label{eqn:wcrossnu}
\ee
where the integral over $\nu$ ranges from $-\infty$ to $+\infty$
(since both $r_1$ and $r_2$ vary between 0 to $+\infty$).
Now,
\be
\int \frac{dx}{\left(a^2 + x^2\right)^{\gamma/2}} = 
a^{1-\gamma}\int \frac{dx}{\left(1 + x^2\right)^{\gamma/2}}
\ee
Therefore, equation~(\ref{eqn:wcrossnu}) becomes
\be
w_{c}(f_{1},f_{2},f_3,f_4,\theta) =
r_0^{\gamma}H_{\gamma}\theta^{1-\gamma}
\left(\frac{f_4}{f_2}\right)^{3/2}
\frac{
\int r^{5-\gamma}dr\psi_1(r)\psi_1(rf_4/f_2)
}
{
\left[ \int r^2dr\psi_1(r) \right]^2
}
\label{eqn:wcross5mg}
\ee
where 
\be
H_{\gamma} = \int_{-\infty}^{+\infty} \frac{dx}{(1+x^2)^{\gamma/2}}
= \frac{(-1/2)!(\gamma/2 - 3/2)!}{(\gamma/2-1)!}
\ee
Since $w_{a}(f_1,f_2,\theta) = w_{c}(f_1,f_2,f_1,f_2,\theta)$, 
equation~(\ref{eqn:wcross5mg}) can be written as
\be
w_{c}(f_{1},f_{2},f_3,f_4,\theta) =
w_{a}(f_1,f_2,\theta)\left(\frac{f_4}{f_2}\right)^{3/2}
\frac{
\int r^{5-\gamma}\psi_1(r)\psi_1(r\sqrt{f_4/f_2})dr
}
{
\int r^{5-\gamma}\left[\psi_1(r)\right]^2dr
}
\label{eqn:wcrossauto}
\ee
We use this relation between 
$w_{c}(f_{1},f_{2},f_3,f_4,\theta) $ and 
$w_{a}(f_1,f_2,\theta)$ to construct the covariance matrix 
of the galaxy counts in different magnitude bins.
We evaluate the integrals in equation~(\ref{eqn:wcrossauto})
using the luminosity function in the appropriate SDSS passband derived by 
Blanton et al. (2000).

\clearpage

\clearpage

\clearpage
\begin{figure}
\plottwo{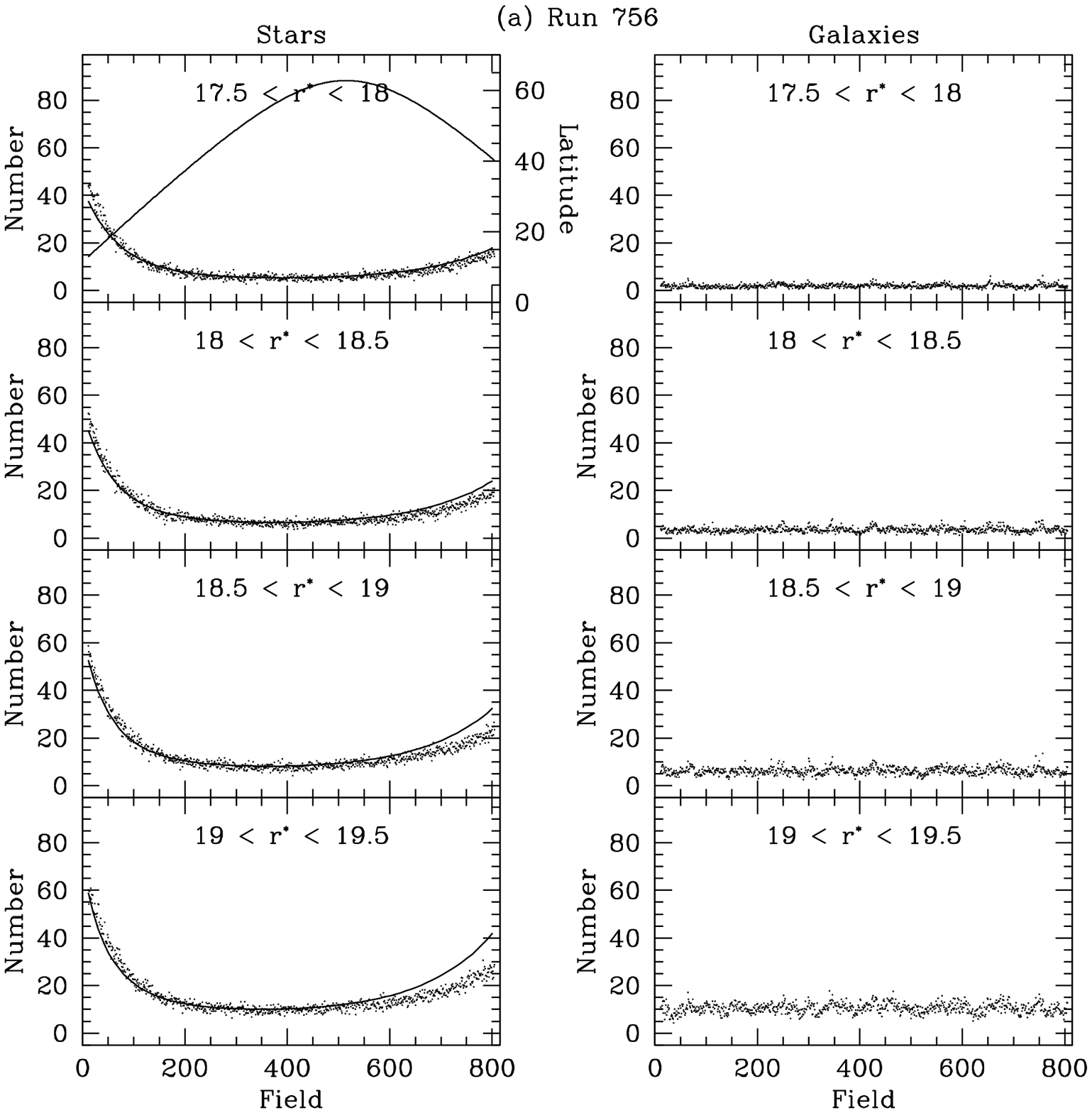}{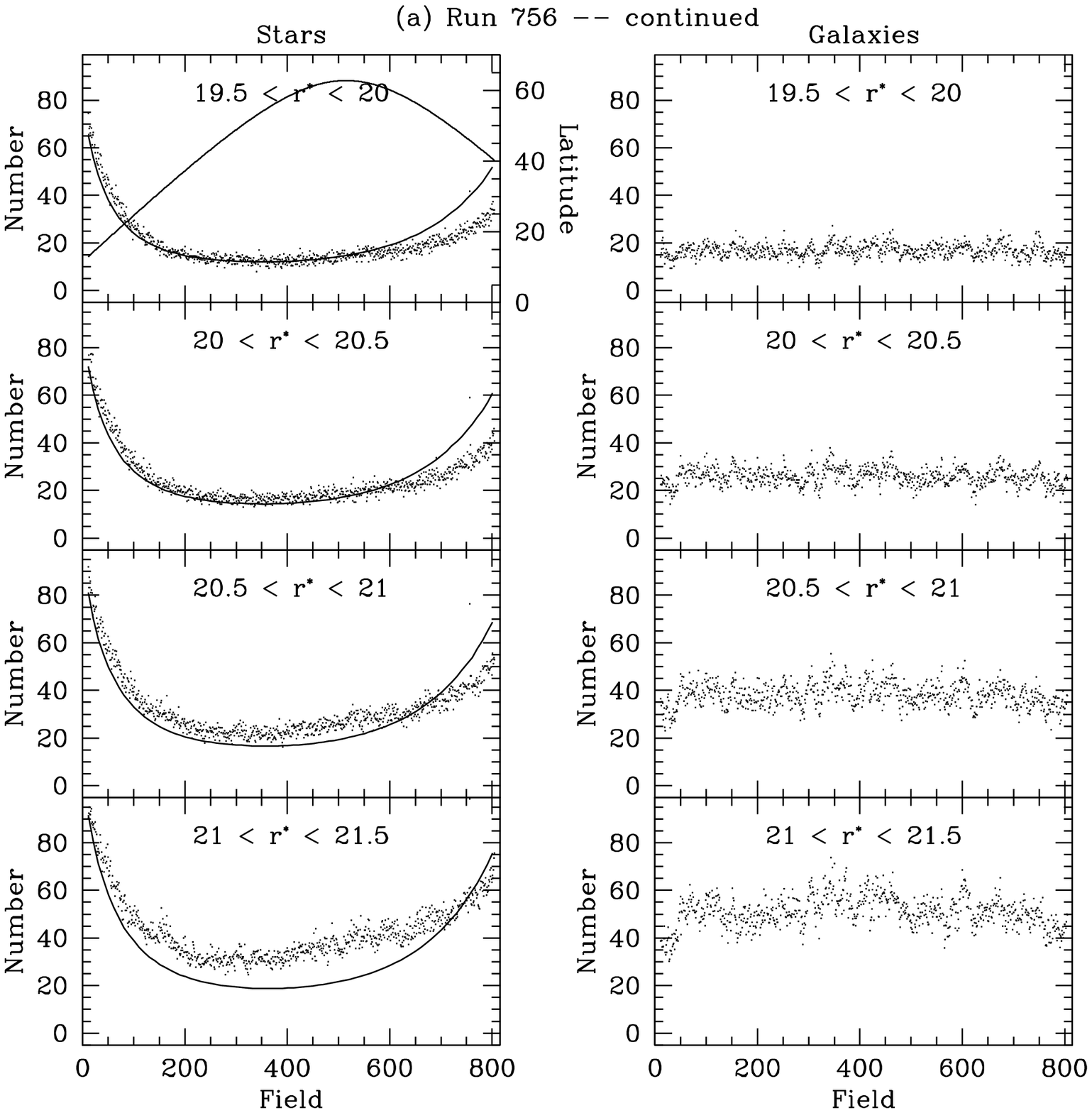}
\caption{Field to field variation of the number of stars and galaxies in
(a) Run 756 and (b) Run 752, averaged over all six CCD columns.
The solid curves in the left panels show the variation of star counts
predicted by the Bahcall-Soneira model.
The top left panel also shows the variation of Galactic latitude over
the duration of this run.}
\label{field_by_field_counts}
\end{figure}

\addtocounter{figure}{-1}
\begin{figure}
\plottwo{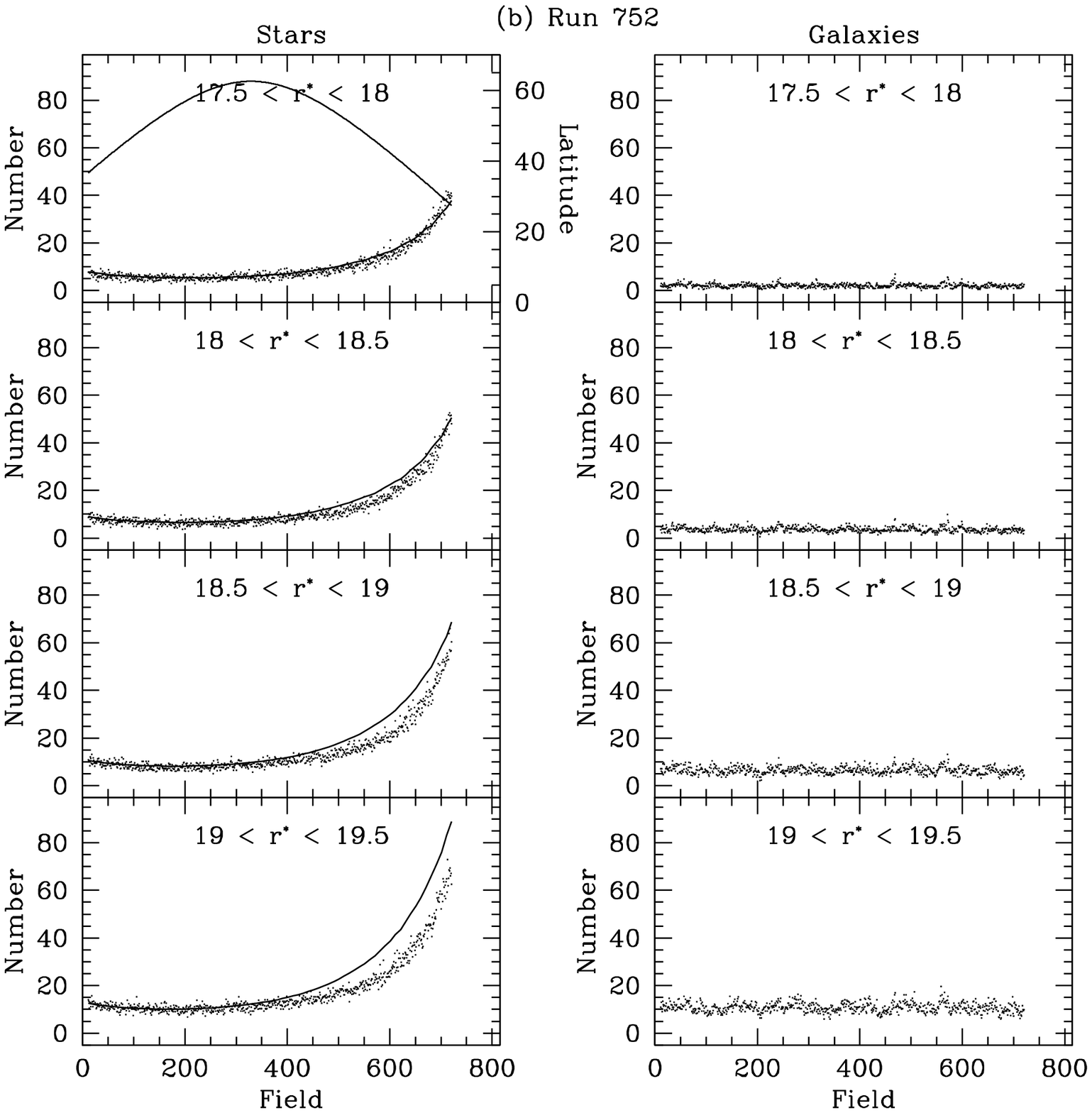}{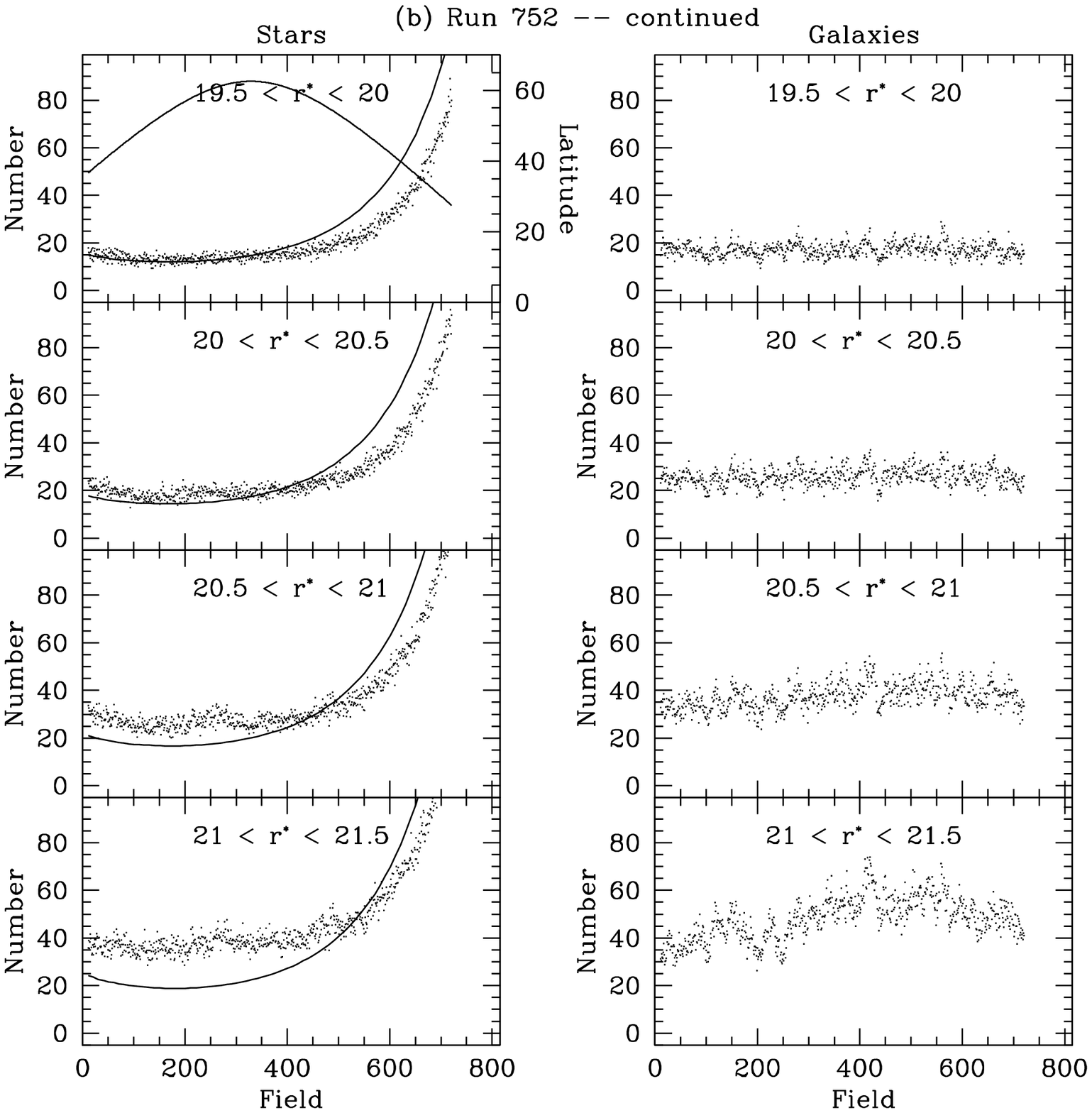}
\caption{continued}
\end{figure}

\clearpage
\begin{figure}
\plotone{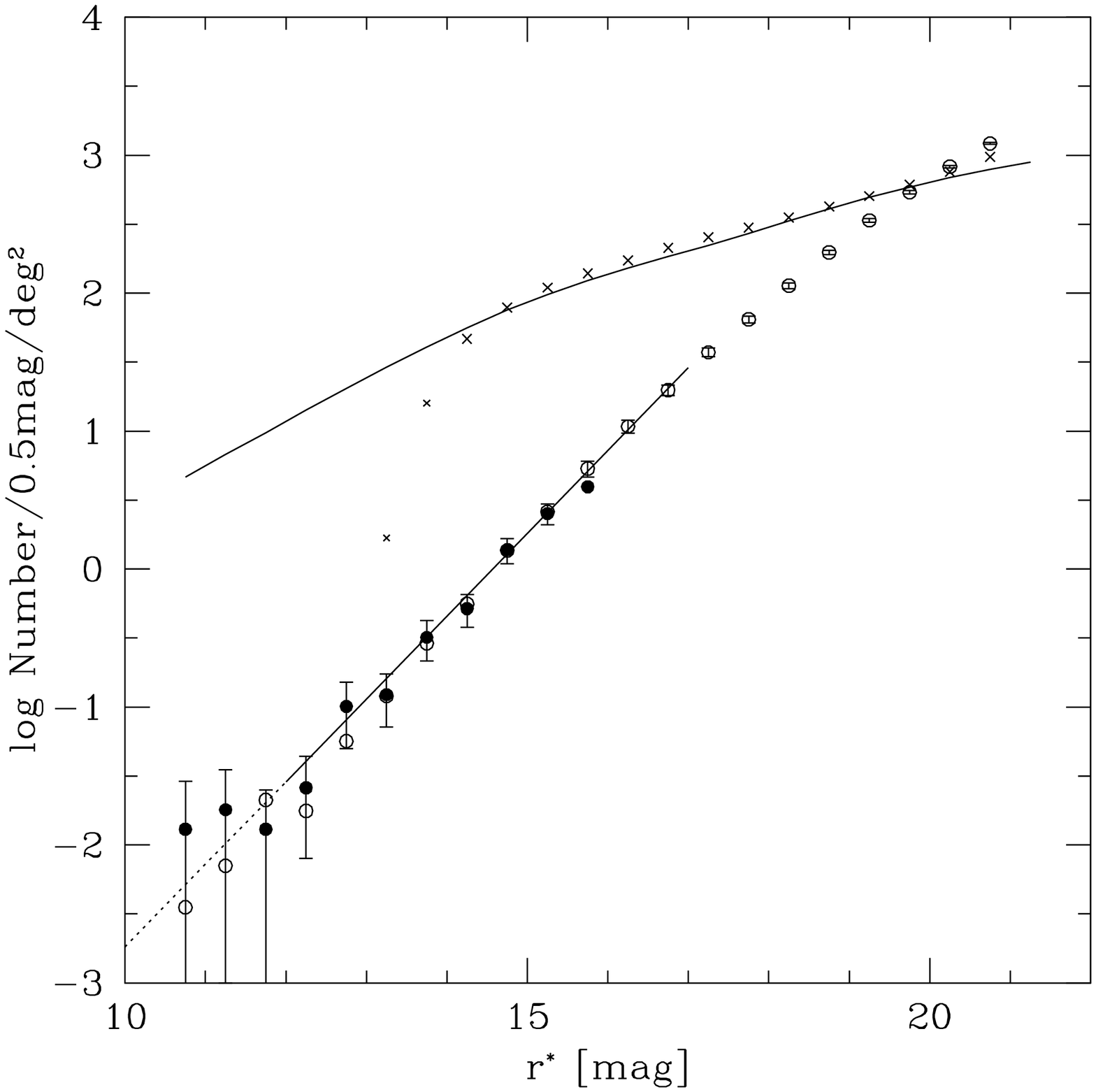}
\caption{Number counts of galaxies as a function of magnitude in the
$r'$ band. Solid points show the galaxy counts from 
the visually inspected sample, and open points from the machine-selected 
sample. The error bars include contributions from both shot-noise
and large scale structure (see text for details).
The line segment shows the counts-magnitude relation expected for a
homogeneous galaxy distribution in a universe with ``Euclidean'' geometry: 
$N(r^*) = A_{r^*}10^{0.6r^*}$. The crosses show the observed star counts
(small crosses at $r^* < 14.5$ show the data where stars
saturate in the SDSS images, and therefore suffer from
incompleteness). The solid curve shows
the prediction of the Bahcall-Soneira model.}
\label{r_counts}
\end{figure}

\clearpage
\begin{figure}
\plotone{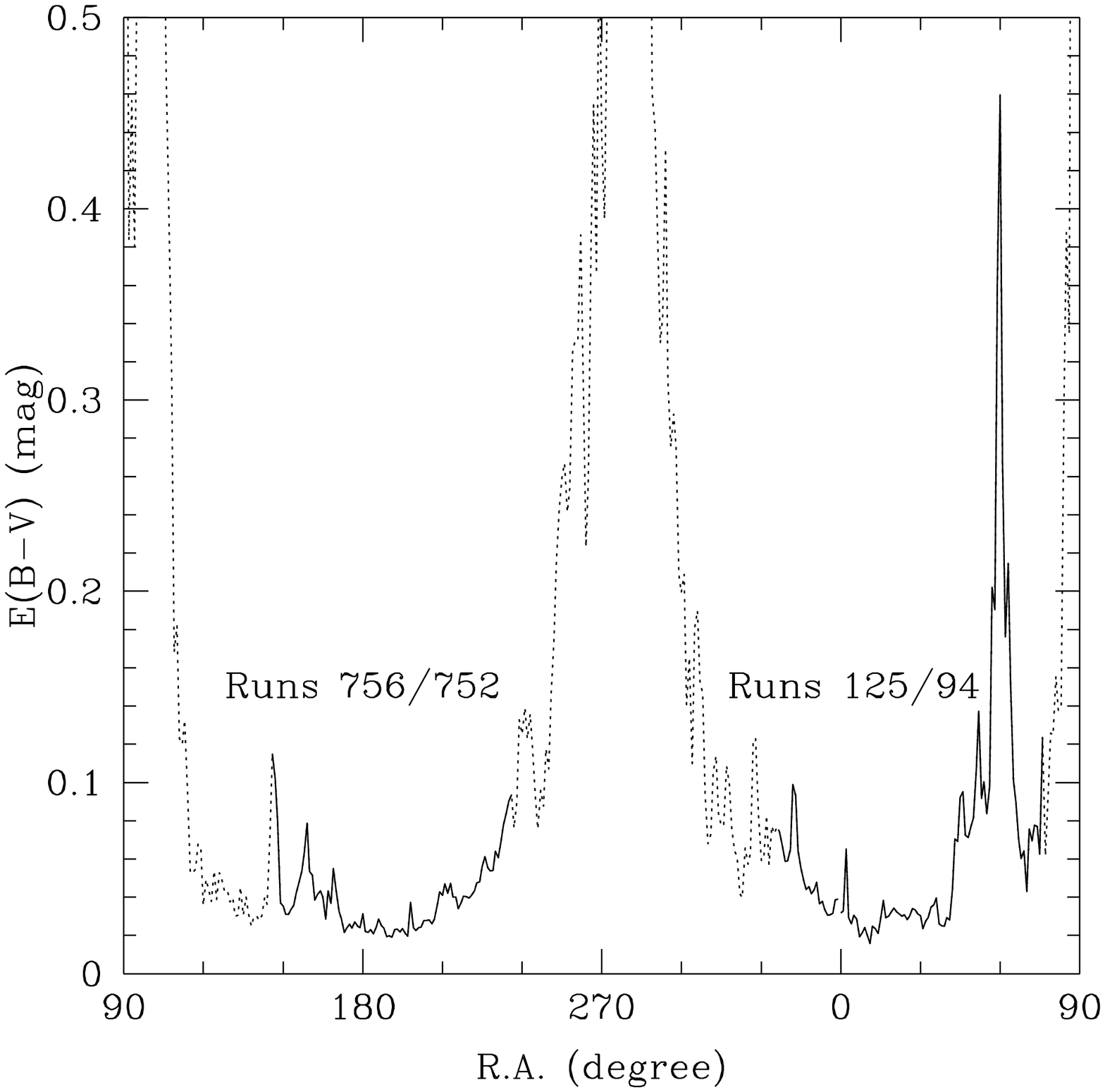}
\caption{Selective extinction $E(B-V)$ from \citet{SchlegelFinkbeinerDavis}
along the Celestial Equator. The solid curves show the 
ranges included in Runs 752/756 (North) and Runs 94/125 (South).}
\label{extinction}
\end{figure}

\clearpage
\begin{figure}
\plotone{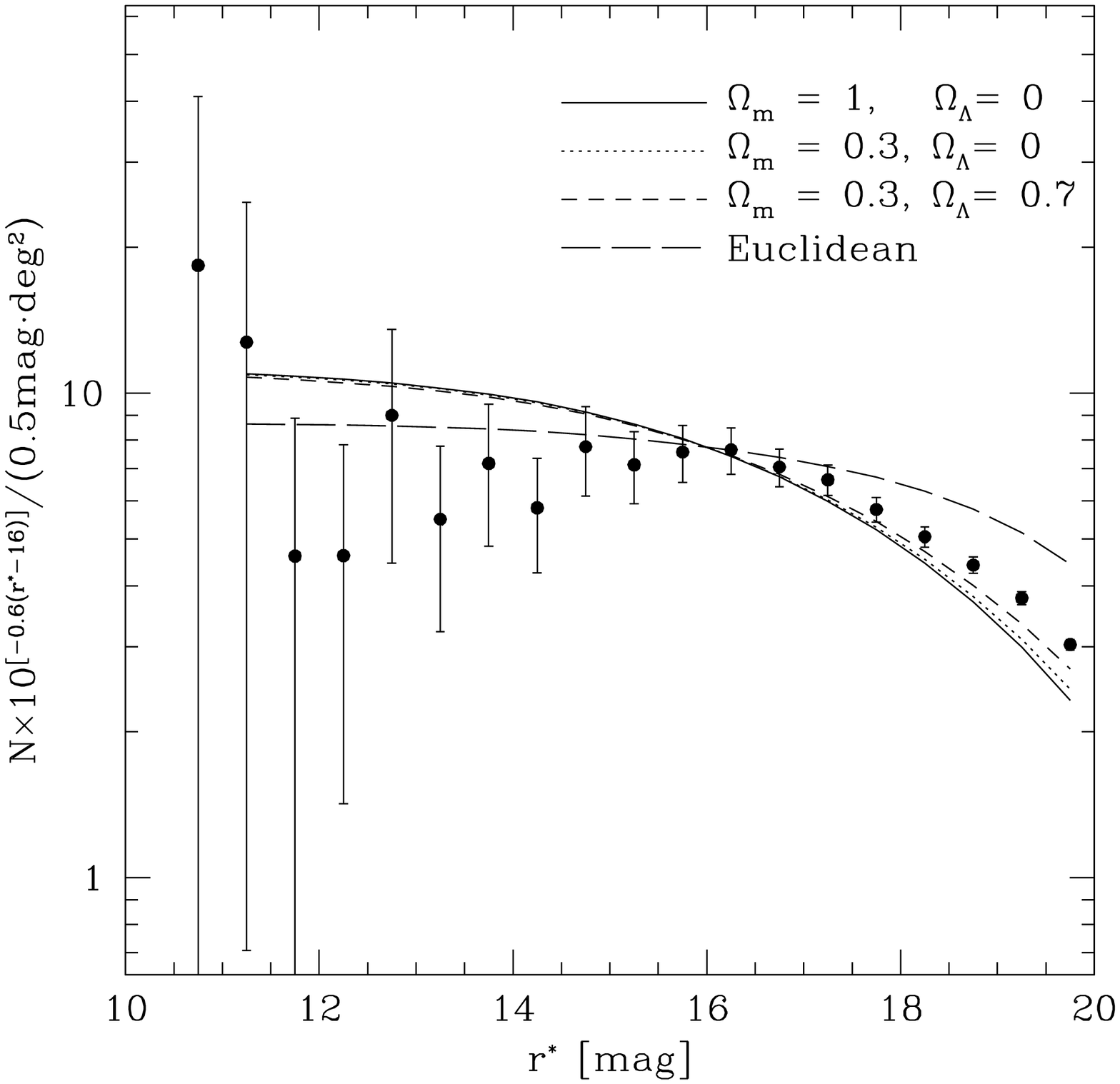}
\caption{Galaxy counts-magnitude relation in the $r^*$ band,
normalized by the expected growth rate in an ``Euclidean'' universe
i.e. $N(r^*)\times10^{[-0.6(r^*-16)]}$.
Points show the observed galaxy counts, while the curves show 
the predictions of a no-evolution model in three different
cosmologies.
All the model curves are normalized to the amplitude $\tilde A_{r^*}$
derived using equation~(\ref{eqn:8}).}
\label{r_growth}
\end{figure}

\clearpage
\begin{figure}
\plotone{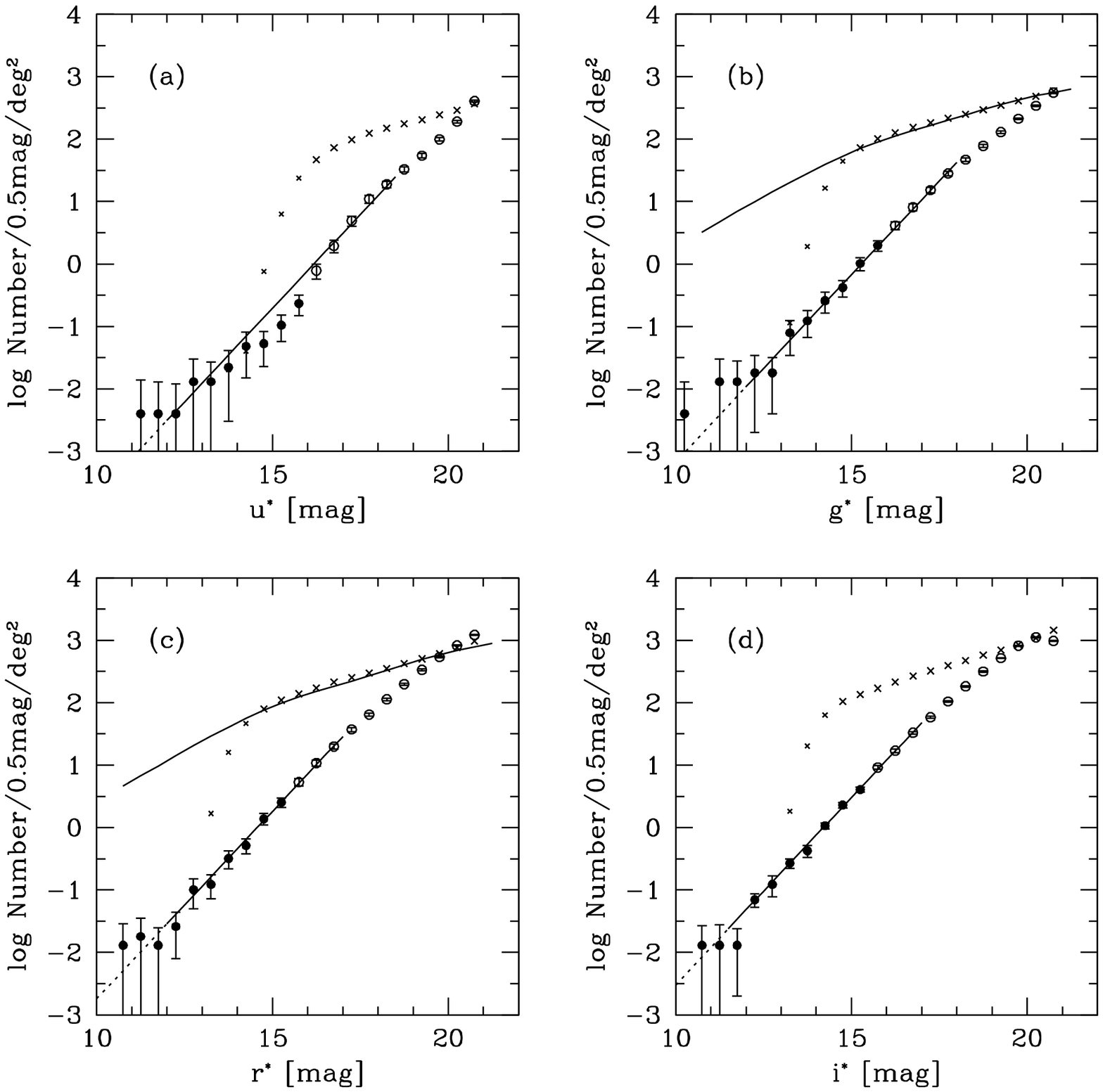}
\caption{Number counts of galaxies as a function of magnitude in the
five color bands $u',\ g',\ r', i'\ $ and $z'$ for the Northern 
equatorial stripe.
Solid points show the
galaxy counts from the visually inspected sample, and open points from
the machine-selected sample. The error bars include contributions from
both shot-noise and large scale structure (see text for details).
The line segment shows the counts-magnitude relation expected in a
homogeneous universe with ``Euclidean'' geometry: 
$N(m) = A_{m}10^{0.6m}$. The crosses show the observed star counts
(small crosses show the data where stars
saturate in the image, and therefore suffer from
incompleteness),
and the solid curve shows the prediction of the Bahcall-Soneira model.}
\label{all_counts}
\end{figure}

\addtocounter{figure}{-1}
\begin{figure}
\plotone{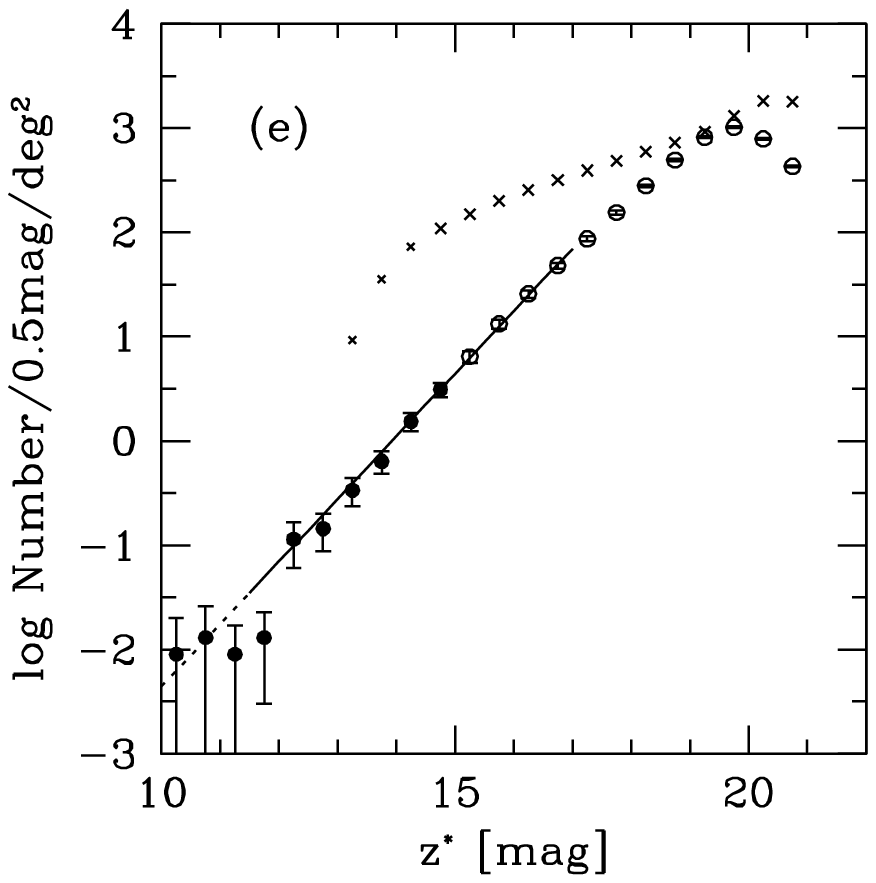}
\caption{continued}
\end{figure}

\clearpage
\begin{figure}
\plotone{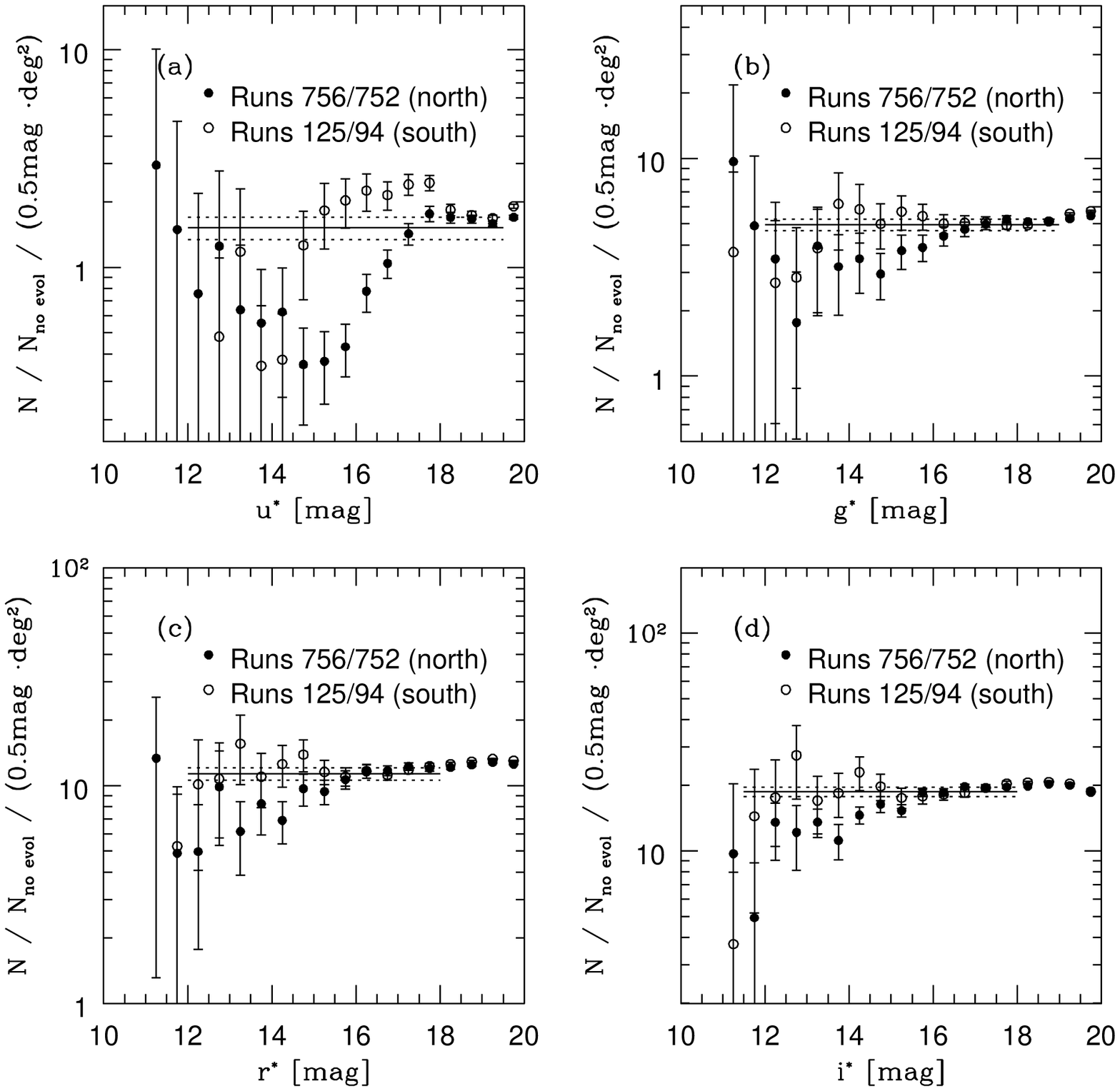}
\caption{Galaxy counts-magnitude relation in the 
$u',\ g',\ r',\ i'\ z'$, and $B$ 
bands, normalized by the no-evolution model prediction
$N(r')_{\rm no~evol}$.
Solid points show the observed galaxy counts from the Northern Equatorial
stripe, and open points are those from the Southern Equatorial stripe 
as discussed in \S8.
The solid lines correspond to $\tilde A_{\lambda^*}$ and dotted lines show
one sigma error ranges.}
\label{Afitting}
\end{figure}

\addtocounter{figure}{-1}
\begin{figure}
\plotone{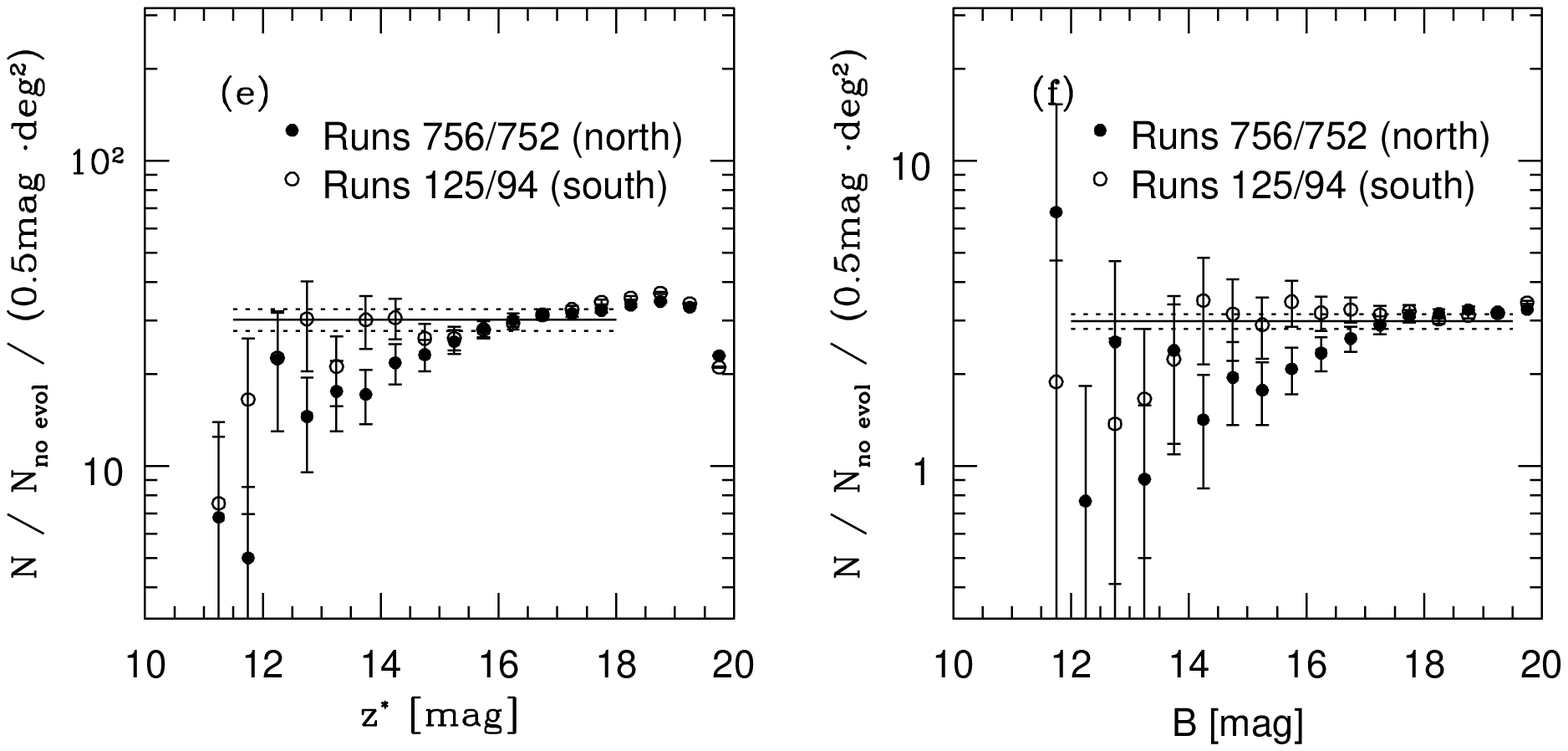}
\caption{continued}
\end{figure}

\clearpage
\begin{figure}
\plotone{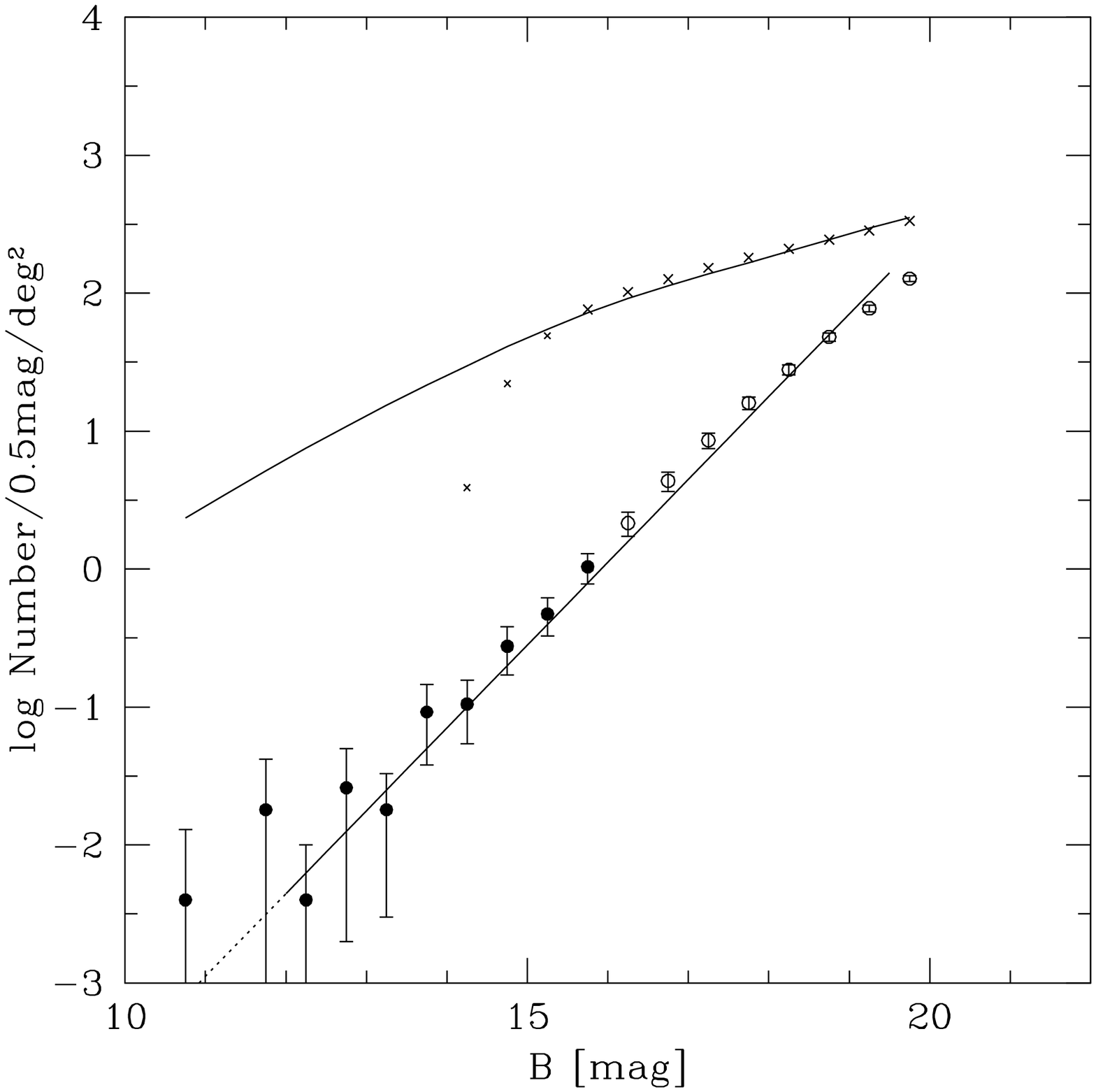}
\caption{Number counts of galaxies as a function of magnitude in the
$B$ band, using the color transformation given in equation~(\ref{eqn:9}).
Solid points show the galaxy counts from 
the visually inspected sample, and open points from the machine-selected 
sample. The error bars include contributions from both shot-noise
and large scale structure (see text for details).
The line segment shows the counts-magnitude relation expected in a
homogeneous universe with ``Euclidean'' geometry: 
$N(B) = A_{B}10^{0.6B}$. The crosses show the observed star counts
(small crosses show the data where stars
saturate in the image, and therefore suffer from
incompleteness), and
the solid curve shows the prediction of the Bahcall-Soneira model.}
\label{B_counts}
\end{figure}

\clearpage
\begin{figure}
\plotone{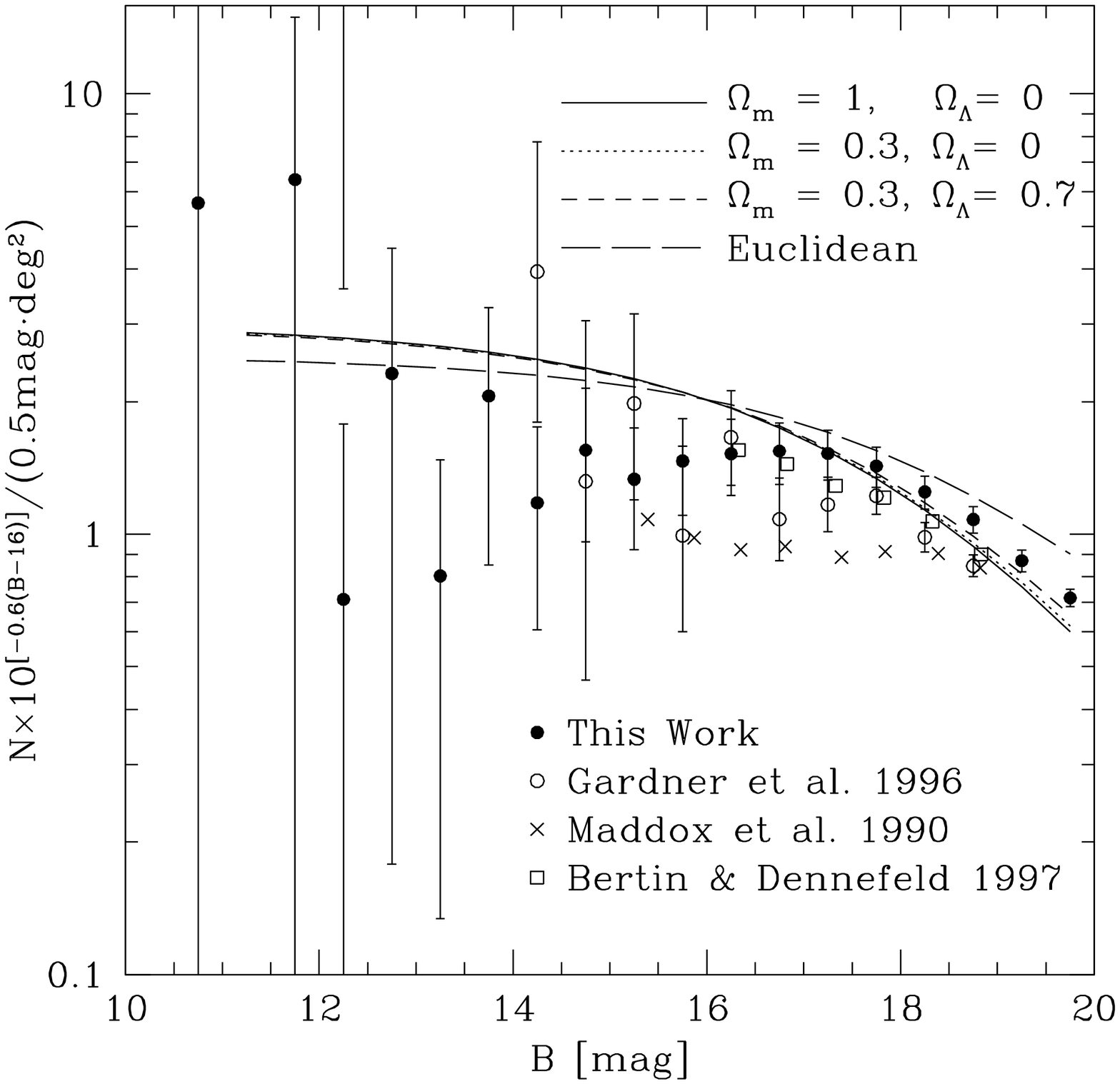}
\caption{Galaxy counts-magnitude relation in the $B$ band,
normalized by the expected growth rate in an ``Euclidean'' universe
i.e. $N(B)\times10^{[-0.6(B-16)]}$.
Solid points show the observed galaxy counts from this work, while the 
curves show the predictions of a no-evolution model in three different
cosmologies.
All the model curves are normalized to the amplitude $\tilde A_{B}$ of the
data. The open circles, crosses and open squares show the 
data on galaxy counts in $B$ band from three previous studies.}
\label{B_growth}
\end{figure}

\clearpage
\begin{figure}
\plotone{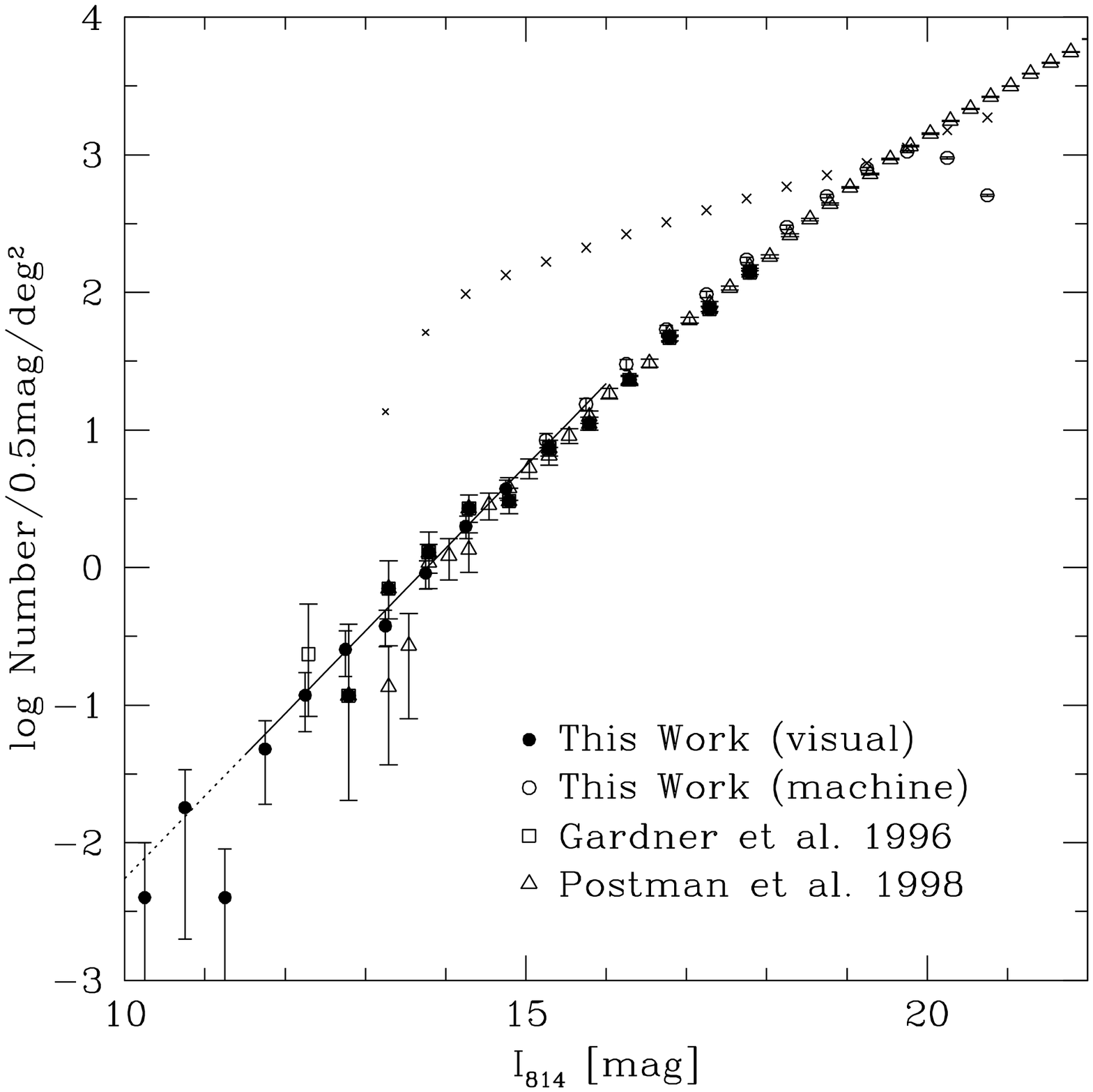}
\caption{Number counts of galaxies as a function of magnitude in the
$I_{814}$ band, using the color transformation given in 
equation~(\ref{eqn:10}) .
Solid circles show the galaxy counts from the visually inspected
sample, and open circles from the machine-selected 
sample. The error bars include contributions from both shot-noise
and large scale structure.
The line segment shows the counts-magnitude relation expected in a
homogeneous universe with ``Euclidean'' geometry: 
$N(I) = A_{I}10^{0.6I}$. The crosses show the observed star counts.
The squares and triangles show galaxy counts from
previous studies.}
\label{I_counts}
\end{figure}

\clearpage
\begin{figure}
\plotone{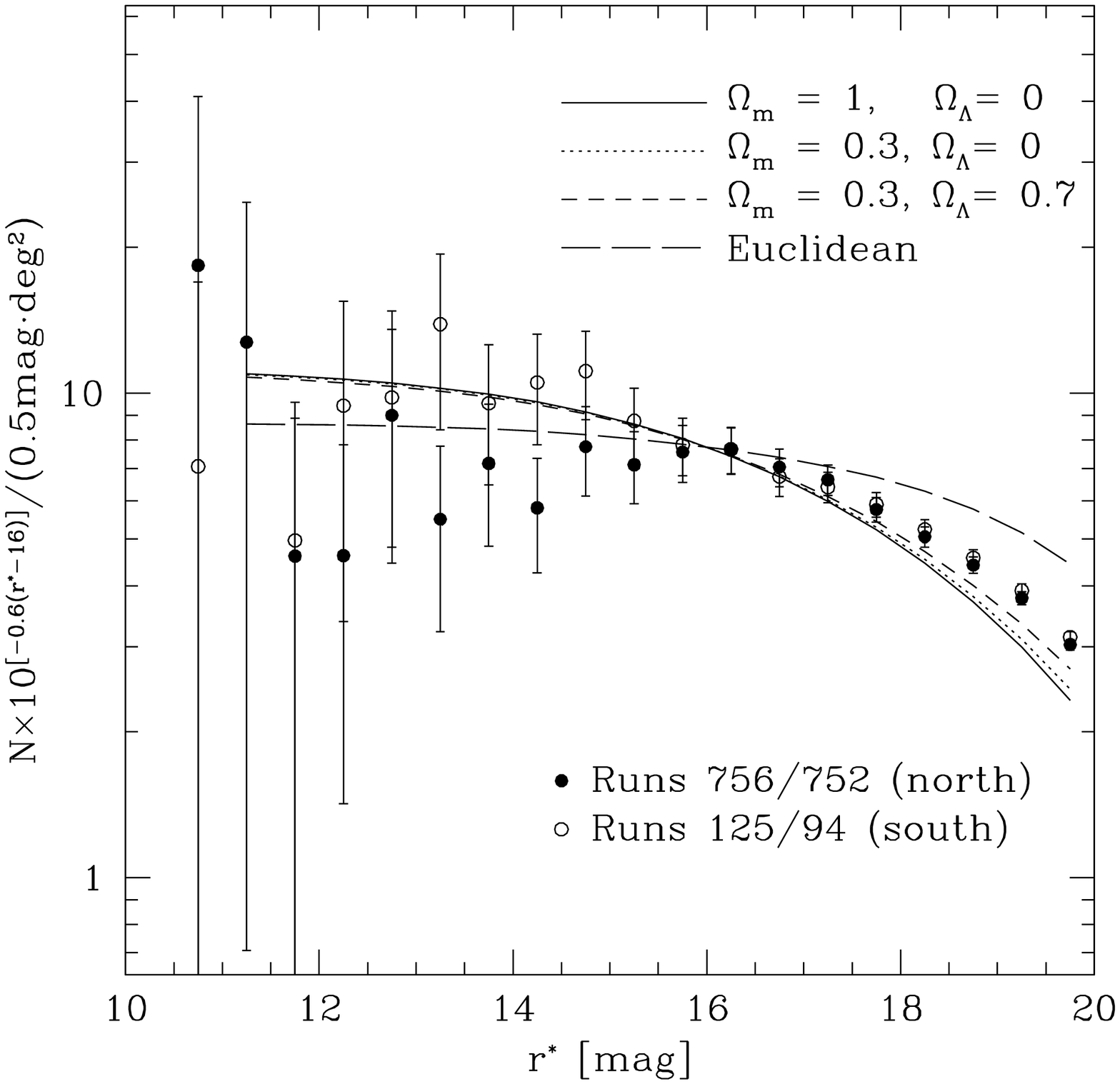}
\caption{Galaxy counts-magnitude relation in the $r^*$ band,
normalized by the expected growth rate in an ``Euclidean'' universe
i.e. $N(r^*)\times10^{[-0.6(r^*-16)]}$.
Solid points show the galaxy counts from the Northern equatorial
stripe, and open points show the counts from the Southern equatorial stripe.
The curves show the predictions of a no-evolution model in three different
cosmologies.
The model curves are normalized to the amplitude $A_{r^*}$ of the
data from the Northern equatorial stripe. }
\label{r_growth_with_south}
\end{figure}

\clearpage
\begin{figure}
\plotone{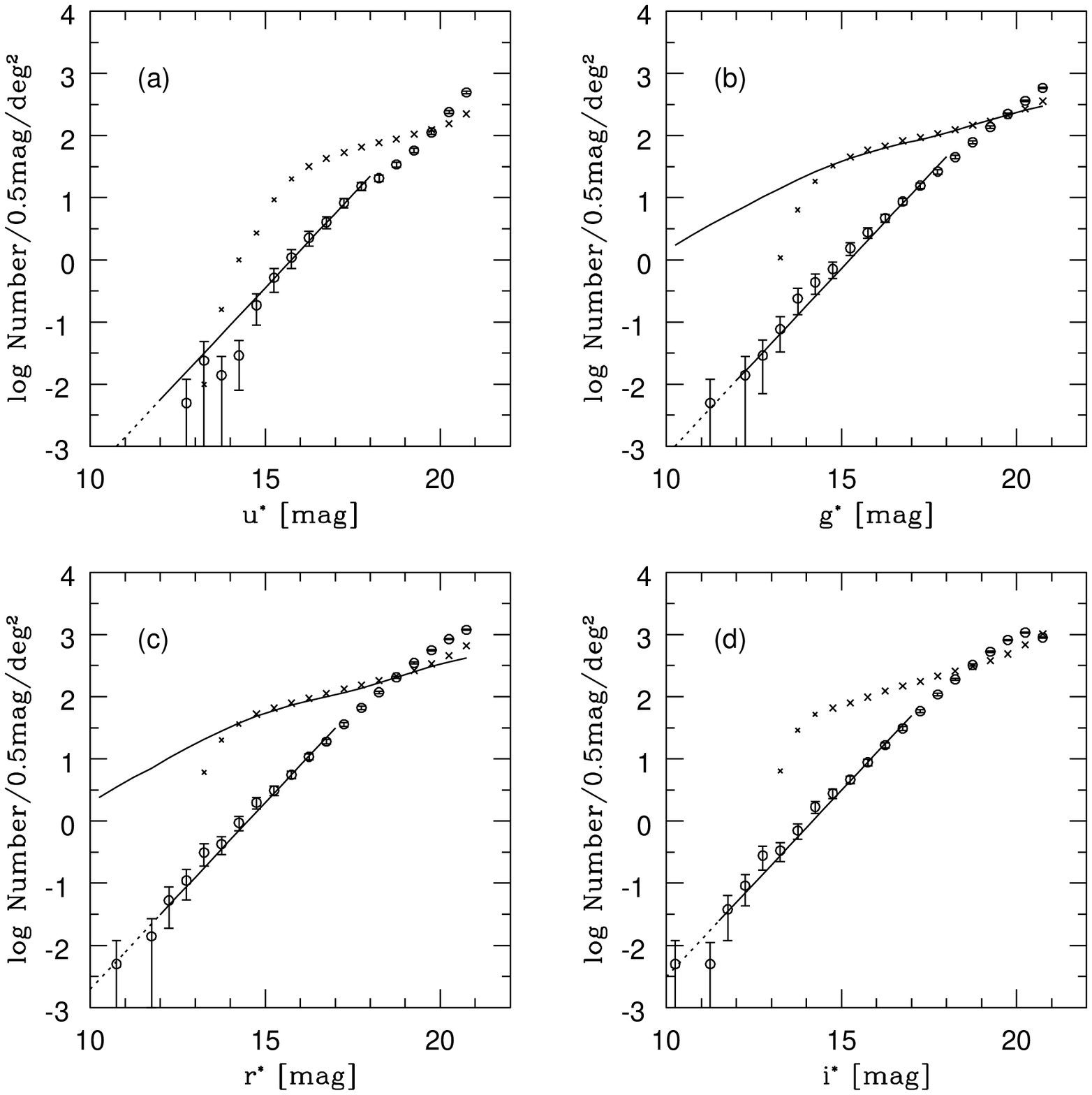}
\caption{Number counts of galaxies as a function of magnitude in the
five color bands $u',\ g',\ r', i'\ $ and $z'$ for the Southern
equatorial stripe.
Open points show the galaxy counts from
the machine-selected sample. The error bars include contributions from
both shot-noise and large scale structure (see text for details).
The line segment shows the counts-magnitude relation expected in a
homogeneous universe with ``Euclidean'' geometry: 
$N(m) = A_{m}10^{0.6m}$. The crosses show the observed star counts
(small crosses show the data where stars
saturate in the image, and therefore suffer from
incompleteness),
and the solid curve shows the prediction of the Bahcall-Soneira model.}
\label{all_counts_south}
\end{figure}

\addtocounter{figure}{-1}
\begin{figure}
\plotone{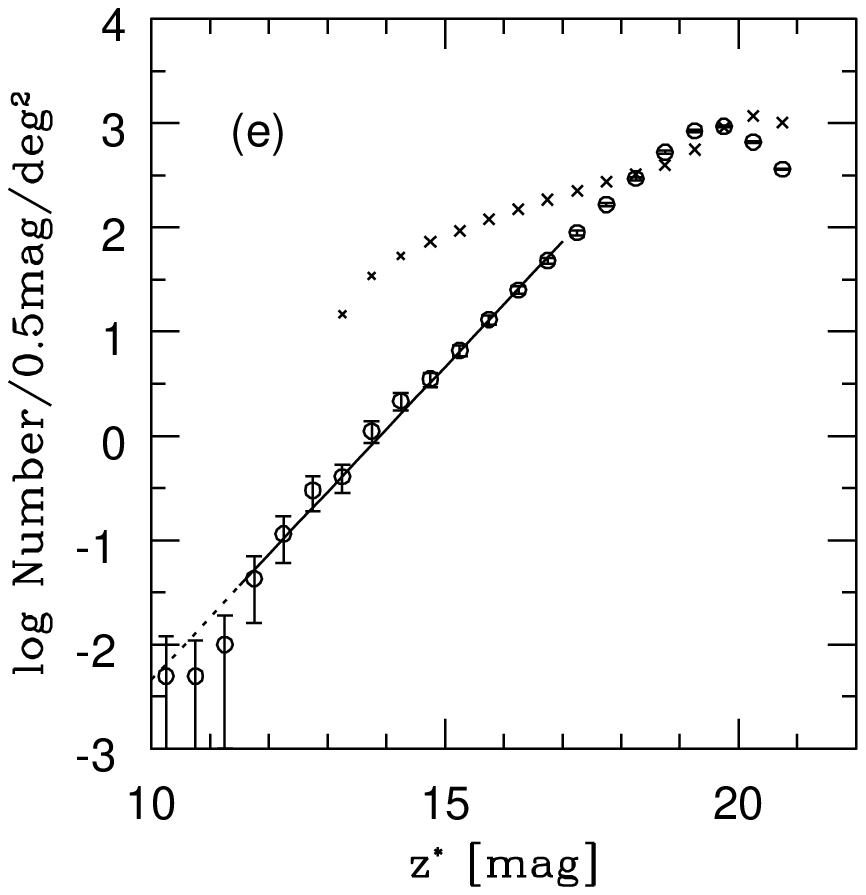}
\caption{continued}
\end{figure}

\clearpage
\begin{figure}
\plotone{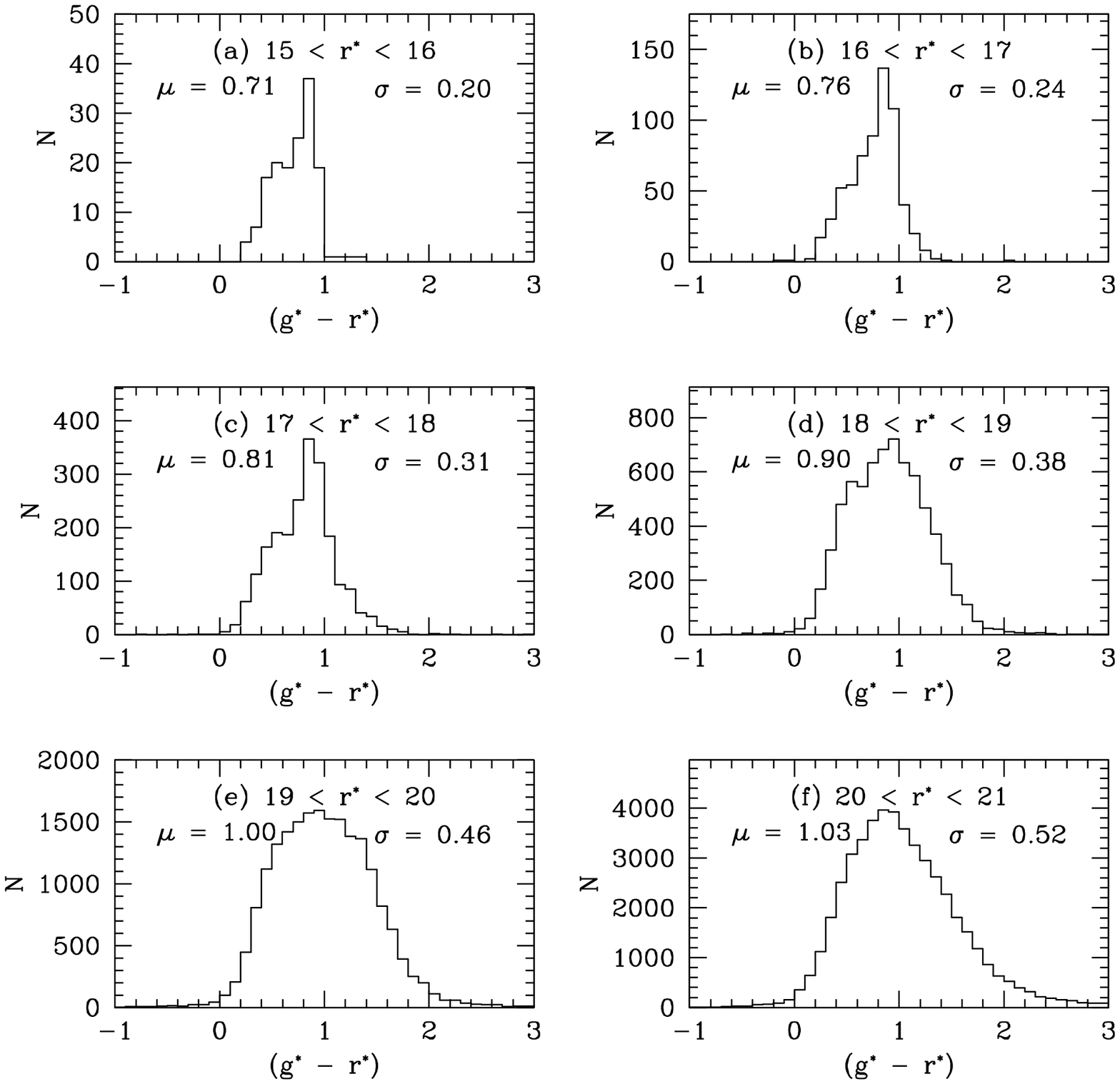}
\caption{Distribution of $g^*-r^*$ colors of all galaxies in the
third CCD column of Run 756. The colors are computed from the
dereddened Petrosian magnitudes of galaxies. Panels (a) to (f) show this 
distribution in different bins of $r'$ Petrosian magnitude. The
mean $(\mu)$ and scatter ($\sigma)$ of the distribution are listed in
each panel.}
\label{hist_g_r}
\end{figure}

\clearpage
\begin{figure}
\plotone{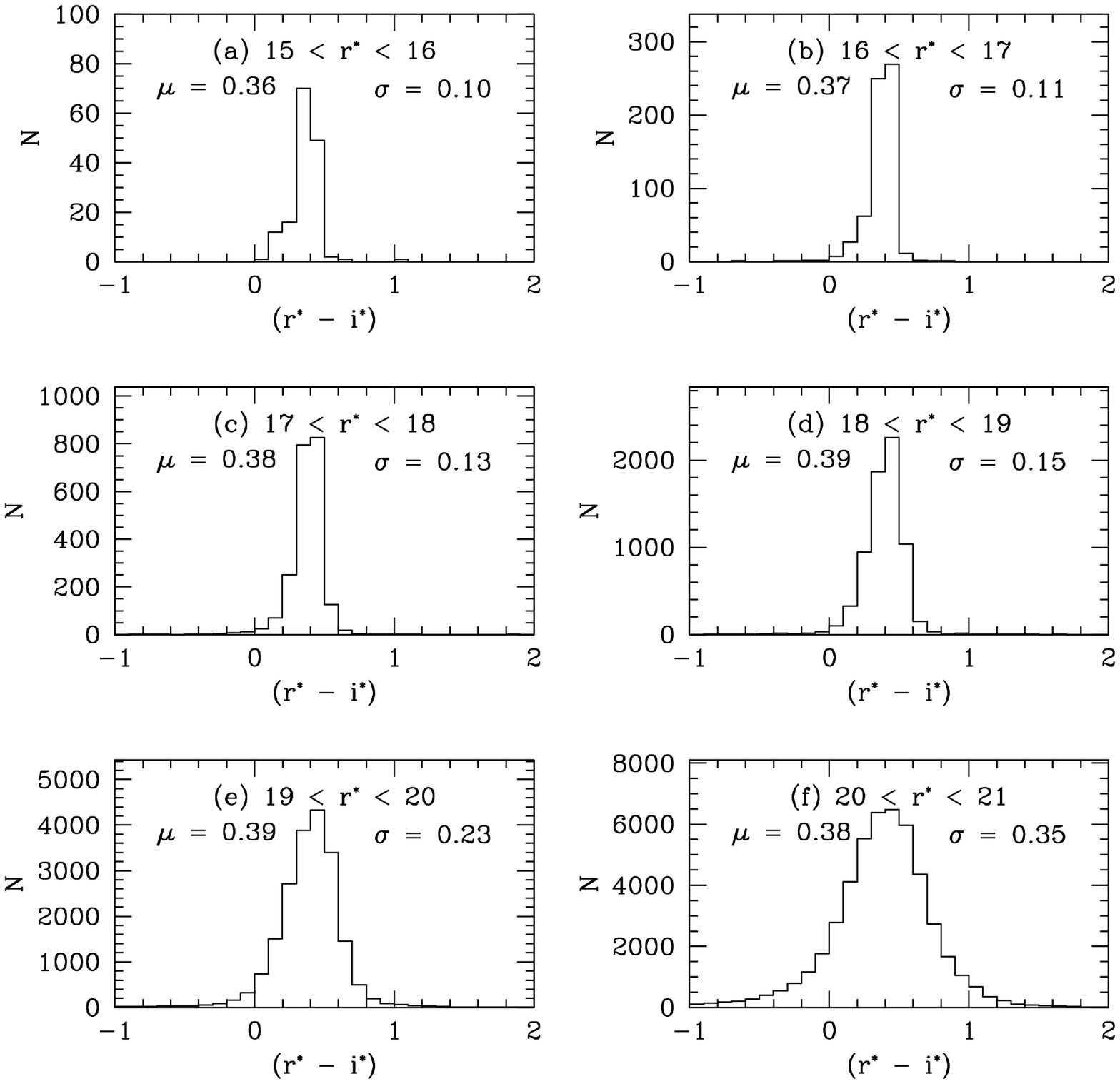}
\caption{Distribution of $r^*-i^*$ colors of all galaxies in the
third CCD column of Run 756. The colors are computed from the
dereddened Petrosian magnitudes of galaxies. Panels (a) to (f) show this 
distribution in different bins of $r^*$ Petrosian magnitude. The
mean $(\mu)$ and scatter ($\sigma)$ of the distribution are listed in
each panel.}
\label{hist_r_i}
\end{figure}

\clearpage
\begin{deluxetable}{lcccc}
\tablewidth{0pc}
\tablecaption{Comparison of the visual and machine galaxy catalogs at
bright magnitudes.
\label{table1}}
\tablehead{\colhead{Magnitude range} & \colhead{Visual counts} 
& \colhead{Machine counts} & \colhead{Visual only} & \colhead{Machine only}}
\startdata
9.5-10.0  &    1 &    0 &  1  &  0 \cr
10.0-10.5 &    0 &    0 &  0  &  0 \cr
10.5-11.0 &    1 &    1 &  0 &   0 \cr
11.0-11.5 &    4 &    1 &  3 &   0 \cr
11.5-12.0 &    4 &    5 &  0 &   1 \cr
12.0-12.5 &    4 &    3 &  1 &   0 \cr
12.5-13.0 &   10 &   14 &  0 &   4 \cr
13.0-13.5 &   30 &   32 &  3 &   5 \cr
13.5-14.0 &   58 &   75 &  4 &  21 \cr
14.0-14.5 &  101 &  114 & 12 &  25 \cr
14.5-15.0 &  257 &  279 & 15 &  37 \cr
15.0-15.5 &  517 &  548 & 24 &  55 \cr
15.5-16.0 &  987 & 1024 & 30 &  67 \cr
          &      &      &    &     \cr
Total     & 1974 & 2096 & 93 & 215 \cr
\enddata
\end{deluxetable}

\clearpage
\begin{deluxetable}{lccccccccccc}
\tabletypesize{\scriptsize}
\tablewidth{0pc}
\tablecaption{Number counts of galaxies in the $u'$, $g'$, $r'$, $i'$, and 
$z'$ color 
bands for the Northern equatorial stripe.  ${\cal N}$ is the number of
galaxies in the sample, and $N$ is the normalized counts
in units of counts per 0.5 mag per deg$^{2}$. Results flagged with a
`v' are based on the visual galaxy catalog (\S~\ref{sec:galcatalog}),
while the rest are based on the machine galaxy catalog.
The galaxy counts at magnitudes fainter than 18 are based on Run 
756 alone.
 \label{table2}}
\tablehead{\colhead{Mag range} & \colhead{Area} & 
\colhead{${\cal N}_{u^*}$} & \colhead{$N_{u^*}$} & 
\colhead{${\cal N}_{g^*}$} & \colhead{$N_{g^*}$} & 
\colhead{${\cal N}_{r^*}$} & \colhead{$N_{r^*}$} & 
\colhead{${\cal N}_{i^*}$} & \colhead{$N_{i^*}$} & 
\colhead{${\cal N}_{z^*}$} & \colhead{$N_{z^*}$} \cr
(mag) & (deg$^2$) & & & & & & & & & &}
\startdata
11.0-11.5 & 228v &      1 &    0.004 &      3 &    0.013 &      4 &    0.018 &      3 &    0.013 &      2 &    0.009 \cr
11.5-12.0 & 228v &      1 &    0.004 &      3 &    0.013 &      3 &    0.013 &      3 &    0.013 &      3 &    0.013 \cr
12.0-12.5 & 228v &      1 &    0.004 &      4 &    0.018 &      6 &    0.026 &     16 &    0.070 &     26 &    0.114 \cr
12.5-13.0 & 228v &      3 &    0.013 &      4 &    0.018 &     23 &    0.101 &     28 &    0.123 &     33 &    0.144 \cr
13.0-13.5 & 228v &      3 &    0.013 &     18 &    0.079 &     28 &    0.123 &     61 &    0.267 &     77 &    0.337 \cr
13.5-14.0 & 228v &      5 &    0.022 &     28 &    0.123 &     73 &    0.320 &     97 &    0.425 &    145 &    0.635 \cr
14.0-14.5 & 228v &     11 &    0.048 &     59 &    0.258 &    118 &    0.517 &    244 &    1.07  &    352 &    1.54  \cr
14.5-15.0 & 228v &     12 &    0.053 &     96 &    0.420 &    315 &    1.38  &    523 &    2.29  &    710 &    3.11  \cr
15.0-15.5 & 228v &     24 &    0.105 &    233 &    1.02  &    577 &    2.53  &    924 &    4.05  &        &          \cr
          & 283  &        &          & 	  & 	     & 	      & 	 & 	  &          &   1819 &    6.44  \cr
15.5-16.0 & 228v &     53 &    0.232 &    452 &    1.98  &        &          &        &          &        &          \cr
          & 283  &        &          &        &          &   1513 &    5.35  &   2597 &    9.20  &   3734 &   13.2   \cr
16.0-16.5 & 283  &    222 &    0.784 &   1170 &    4.14  &   3050 &   10.8   &   4806 &   17.0   &   7268 &   25.7   \cr
16.5-17.0 & 283  &    555 &    1.96  &   2290 &    8.1   &   5607 &   19.8   &   9337 &   33.0   &  13601 &   48.1   \cr
17.0-17.5 & 283  &   1393 &    4.92  &   4312 &   15.3   &  10532 &   37.3   &  16474 &   58.3   &  24426 &   86.4   \cr
17.5-18.0 & 283  &   3121 &   11.0   &   7939 &   28.1   &  18239 &   64.5   &  29553 &  104.6   &  43837 &  155.1   \cr
18.0-18.5 & 149  &   2823 &   18.9   &   6976 &   46.8   &  16890 &  113.2   &  27232 &  182.5   &  41673 &  279.4   \cr
18.5-19.0 & 149  &   4898 &   32.9   &  11653 &   78.1   &  29443 &  197.4   &  47046 &  315.4   &  73714 &  494.1   \cr
19.0-19.5 & 149  &   8105 &   54.4   &  19359 &  129.8   &  50206 &  336.5   &  77322 &  518.3   & 121858 &  816.9   \cr
19.5-20.0 & 149  &  14799 &   99.3   &  31606 &  211.9   &  80324 &  538.4   & 121858 &  816.9   & 151616 & 1016     \cr
20.0-20.5 & 149  &  28359 &  190.3   &  51000 &  341.8   & 123213 &  825.9   & 166314 & 1114     & 117358 &  786.7   \cr
20.5-21.0 & 149  &  60618 &  406.8   &  82614 &  553.8   & 181428 & 1216     & 145706 &  976.7   &  64083 &  429.6   \cr
21.0-21.5 & 149  & 111725 &  749.8   & 131641 &  882.4   & 242743 & 1627     &  65955 &  442.1   &  31661 &  212.2   \cr
\enddata
\end{deluxetable}

\clearpage
\begin{deluxetable}{ccc}
\tablewidth{0pc}
\tablecaption{Normalization coefficients of the $10^{0.6m}$ law,
including the cosmological and K-corrections
\label{table3}}
\tablehead{ \colhead{Band} & \colhead{Fitting range} & \colhead{$\tilde A_{\lambda}$} \cr
$(\lambda)$ & (mag) & [(0.5mag) $^{-1}$deg$^{-2}$]}
\startdata
$u^*$ & 12.0-19.5 & $\ 1.52\pm0.18$ \cr
$g^*$ & 12.0-19.0 & $\ 4.95\pm0.29$ \cr
$r^*$ & 12.0-18.0 & $ 11.30\pm0.75$ \cr
$i^*$ & 11.5-18.0 & $ 18.66\pm0.90$ \cr
$z^*$ & 11.5-18.0 & $ 30.17\pm2.40$ \cr 
$B$   & 12.0-19.5 & $\ 2.98\pm0.17$ \cr
\enddata
\end{deluxetable}

\clearpage
\begin{deluxetable}{lccccc}
\tablewidth{0pc}
\tablecaption{Number counts of galaxies in the $B$ and $I_{814}$ 
bands in the Northern equatorial stripe 
(see Table \ref{table2} for explanations) 
\label{table4}}
\tablehead{\colhead{Mag range} & \colhead{Area} & \colhead{${\cal N}_{B}$} & \colhead{$N_{B}$}& \colhead{${\cal N}_{I_{814}}$} & \colhead{$N_{I_{814}}$} \cr
(mag) & (deg$^2$) & & & &}
\startdata
11.0-11.5 & 228v &      0 &    0.000 &      1 &    0.004 \cr
11.5-12.0 & 228v &      4 &    0.018 &     11 &    0.048 \cr
12.0-12.5 & 228v &      1 &    0.004 &     27 &    0.118 \cr
12.5-13.0 & 228v &      6 &    0.026 &     58 &    0.254 \cr
13.0-13.5 & 228v &      4 &    0.018 &     86 &    0.377 \cr
13.5-14.0 & 228v &     21 &    0.092 &    208 &    0.911 \cr
14.0-14.5 & 228v &     24 &    0.105 &    455 &    2.00 \cr
14.5-15.0 & 228v &     63 &    0.276 &    854 &    3.74 \cr
15.0-15.5 & 228v &    108 &    0.473 &        &         \cr
          & 283 &        &          &   2377 &    8.41 \cr
15.5-16.0 & 228v &    237 &   1.04   &        &          \cr
          & 283 &        &          &   4347 &   15.5 \cr
16.0-16.5 & 283 &    608 &    2.15  &   8501 &   30.1 \cr
16.5-17.0 & 283 &   1230 &    4.35  &  15236 &   53.9 \cr
17.0-17.5 & 283 &   2423 &    8.57  &  27316 &   96.7 \cr
17.5-18.0 & 283 &   4526 &   16.0   &  48717 &  172.4 \cr
18.0-18.5 & 149 &   4167 &   27.9   &  44432 &  297.8 \cr
18.5-19.0 & 149 &   7192 &   48.2   &  74565 &  499.8 \cr
19.0-19.5 & 149 &  11569 &   77.5   & 118000 &  791.0 \cr
19.5-20.0 & 149 &  19017 &  127.4   & 157416 &  1055 \cr
\enddata
\end{deluxetable}

\clearpage
\begin{deluxetable}{cccccc}
\tablewidth{0pc}
\tablecaption{Luminosity density of the universe in different bands.
\label{table5}}
\tablehead{ \colhead{Band} & \colhead{$M^*$} & $\alpha_{\lambda}$ & $\phi_{\lambda}^*$ & $M_\odot$ & ${\cal L}_{\lambda}$ \cr
& & & $10^{-2}\ h^{3}{\rm Mpc}^{-3}$ & & $10^{8}\ L_{\odot} h {\rm Mpc}^{-3}$}
\startdata
$u^*$ & $-18.34\pm0.08$ & $-1.35\pm0.09$ & $5.67\pm1.05$ & 6.38 & $6.30\pm0.85$ \cr
$g^*$ & $-20.04\pm0.04$ & $-1.26\pm0.05$ & $1.81\pm0.16$ & 5.06 & $2.53\pm0.20$ \cr
$r^*$ & $-20.83\pm0.03$ & $-1.20\pm0.03$ & $1.40\pm0.11$ & 4.64 & $2.60\pm0.21$ \cr
$i^*$ & $-21.26\pm0.04$ & $-1.25\pm0.04$ & $1.27\pm0.10$ & 4.53 & $3.31\pm0.23$ \cr
$z^*$ & $-21.55\pm0.04$ & $-1.24\pm0.05$ & $1.38\pm0.14$ & 4.52 & $4.60\pm0.43$ \cr
$B$   & $-19.60{+0.20\atop-0.10}$ & $-1.1\pm0.1$  
& $2.05{+0.67\atop-0.30}$ & 5.46 & $2.41{+0.46\atop-0.31}$ \cr
\enddata
\end{deluxetable}

\clearpage
\begin{deluxetable}{lccccccccccc}
\tabletypesize{\scriptsize}
\tablewidth{0pc}
\tablecaption{Number counts of galaxies in the $u'$, $g'$, $r'$, $i'$, 
and $z'$ bands in the Southern equatorial stripe
(see Table \ref{table2} for explanations) 
\label{table6}}
\tablehead{\colhead{Mag range} & \colhead{Area} & 
\colhead{${\cal N}_{u^*}$} & \colhead{$N_{u^*}$} & 
\colhead{${\cal N}_{g^*}$} & \colhead{$N_{g^*}$} & 
\colhead{${\cal N}_{r^*}$} & \colhead{$N_{r^*}$} & 
\colhead{${\cal N}_{i^*}$} & \colhead{$N_{i^*}$} & 
\colhead{${\cal N}_{z^*}$} & \colhead{$N_{z^*}$} \cr
(mag) & (deg$^2$) & & & & & & & & & &}
\startdata
14.0-14.5 & 209 &      6 &    0.029 &     91 &    0.436 &    196 &    0.939 &    352 &   1.686 &    453 &    2.17 \cr
14.5-15.0 & 209 &     39 &    0.187 &    149 &    0.714 &    413 &    1.98  &    578 &   2.77  &    734 &    3.52 \cr
15.0-15.5 & 209 &    108 &    0.517 &    322 &    1.54  &    650 &    3.11  &    968 &   4.64  &   1381 &    6.62 \cr
15.5-16.0 & 209 &    228 &    1.09  &    576 &    2.76  &   1157 &    5.54  &   1842 &   8.83  &   2731 &   13.08 \cr
16.0-16.5 & 209 &    475 &    2.27  &    984 &    4.71  &   2262 &   10.8   &   3477 &  16.65  &   5251 &   25.1  \cr
16.5-17.0 & 209 &    843 &    4.03  &   1811 &    8.68  &   3962 &   18.9   &   6461 &  30.9   &  10063 &   48.2  \cr
17.0-17.5 & 209 &   1733 &    8.29  &   3274 &   15.7   &   7525 &   36.0   &  12245 &  58.7   &  18648 &   89.3  \cr
17.5-18.0 & 209 &   3197 &   15.3   &   5501 &   26.4   &  13823 &   66.2   &  22653 & 108.5   &  34586 &  165.7  \cr
18.0-18.5 & 100 &   2056 &   20.6   &   4502 &   44.8   &  11724 &  116.8   &  18947 & 188.7   &  29486 &  293.7  \cr
18.5-19.0 & 100 &   3426 &   34.3   &   7786 &   77.5   &  20435 &  203.5   &  32309 & 321.8   &  52662 &  524.5  \cr
19.0-19.5 & 100 &   5752 &   57.5   &  13698 &  136.4   &  34895 &  347.5   &  52893 & 526.8   &  84069 &  837.3  \cr
19.5-20.0 & 100 &  11178 &  111.8   &  22262 &  221.7   &  55831 &  556.0   &  81855 & 815.2   &  93168 &  927.9  \cr
\enddata
\end{deluxetable}

\end{document}